\documentclass[lettersize,journal]{IEEEtran}
\usepackage{amsmath,amsfonts}
\usepackage{array}
\usepackage[caption=false,font=normalsize,labelfont=sf,textfont=sf]{subfig}
\usepackage{textcomp}
\usepackage{stfloats}
\usepackage{url}
\usepackage{verbatim}
\usepackage{graphicx}
\usepackage{cite}
\usepackage{mathtools}
\usepackage{pifont}
\usepackage{times}
\usepackage{soul}
\usepackage{booktabs}  
\usepackage{amsthm}
\newtheorem{definition}{Definition} 
\newtheorem{example}{Example}
\newtheorem{theorem}{Theorem}
\newtheorem{lemma}{Lemma}
\usepackage{multirow}
\usepackage{makecell}
\usepackage{amsthm}
\usepackage[ruled,linesnumbered]{algorithm2e}
\usepackage[colorlinks=true, urlcolor=black, linkcolor=blue, citecolor=blue]{hyperref}
\usepackage{orcidlink}
\begin{document}

\title{Efficient Candidate-Free R-S Set Similarity Joins with Filter-and-Verification Trees on MapReduce}

\author{
    Yuhong Feng\orcidlink{0000-0002-7691-5587}, 
    Fangcao Jian\orcidlink{0009-0001-4312-157X}, 
    Yixuan Cao\orcidlink{0009-0006-6241-4251},
    Xiaobin Jian\orcidlink{0009-0006-1907-0991},
    Jia Wang\orcidlink{0009-0001-6284-7069},
    Haiyue Feng\orcidlink{0009-0002-7239-3467},
    and\\ Chunyan Miao\orcidlink{0000-0002-0300-3448}, \IEEEmembership{Fellow,~IEEE}
    \thanks{
    This work is supported by the Shenzhen Science and Technology Foundation (General Program, JCYJ20210324093212034), 2022 Guangdong Province Undergraduate University Quality Engineering Project (Shenzhen University Academic Affairs [2022] No. 7), Science and Technology R\&D Program of Shenzhen (20220810135520002), Guangdong Province Key Laboratory of Popular High Performance Computers 2017B030314073, and Guangdong Province Engineering Center of China-made High Performance Data Computing System. \textit{(Corresponding author: Yuhong Feng)}}
    \thanks{Yuhong Feng, Fangcao Jian, Yixuan Cao, Xiaobin Jian, and Jia Wang are with the College of Computer Science and Software Engineering, Shenzhen University, Shenzhen 518060, China (e-mail: \url{yuhongf@szu.edu.cn}; \url{jianfangcao2023@email.szu.edu.cn}; \url{caoyixuan2019@email.szu.edu.cn}; \url{jianxiaobin2022@email.szu.edu.cn}; \url{wangjia2023@email.szu.edu.cn}).}
    \thanks{Haiyue Feng and Chunyan Miao are with the College of Computing and Data Science, Nanyang Technological University, Singapore 639798 (e-mail: \url{haiyue002@e.ntu.edu.sg}; \url{ascymiao@ntu.edu.sg}).}
}

\markboth{Journal of \LaTeX\ Class Files,~Vol.~14, No.~8, August~2021}%
{Shell \MakeLowercase{\textit{et al.}}: A Sample Article Using IEEEtran.cls for IEEE Journals}


\maketitle

\begin{abstract}
Given two different collections of sets $\mathcal{R}$ and $\mathcal{S}$, the exact R-S set similarity join ({\em R-S Join}) finds all set pairs with similarity no less than a given threshold, which has widespread applications.
Existing algorithms accelerate large-scale R-S Joins using a two-stage {\em filter-and-verification} framework along with the parallel and distributed MapReduce framework, however, they suffer from excessive candidate set pairs ({\em candidates}), 
leading to significant I/O and verification overhead. 
This paper proposes novel candidate-free R-S Join (CF-RS-Join) algorithms that integrate filtering and verification into a single stage through the {\em filter-and-verification tree} ({\em FVT}) and its linear variant ({\em LFVT}). 
First, CF-RS-Join with FVT ({\em CF-RS-Join/FVT}) is proposed to leverage an innovative FVT structure that compresses elements and associated sets in memory, enabling single-stage processing that eliminates candidate generation, enables fast lookups, and reduces database scans. 
Correctness proofs are provided. 
Second, CF-RS-Join with LFVT ({\em CF-RS-Join/LFVT}) is proposed to exploit a more compact Linear FVT, which compresses non-branching paths into single nodes and stores them in linear arrays for optimized traversal. Third, {\em MR-CF-RS-Join/FVT} and {\em MR-CF-RS-Join/LFVT} are proposed to extend our approaches using MapReduce for parallel processing. 
Extensive experiments have been conducted on the proposed algorithms against state-of-the-art (SOTA) baselines in terms of {\em execution time}, {\em scalability}, {\em memory usage}, and {\em disk usage}.
The results show that
MR-CF-RS-Join/LFVT
outperforms the runner-up by up to 
1.37$\times$-15.78$\times$ on 7 real-world datasets.
\end{abstract}

\begin{IEEEkeywords}
Set Similarity Join, Filter-and-Verification, Data
Compression, Parallel and Distributed Computing
\end{IEEEkeywords}

\section{Introduction}
\IEEEPARstart{G}{iven} two different set collections $\mathcal{R}$ and $\mathcal{S}$, the R-S set similarity join (R-S Join) finds set pairs with similarities no less than the threshold. 
R-S Join is a primitive operator in varied knowledge discovery and data mining applications, e.g., data cleaning~\cite{datacleaning2006}, community mining~\cite{dong2011efficient}, process mining~\cite{Kocher2021}, customized recommendation~\cite{das2007google}, near-duplicate detection~\cite{xiao2011efficient}, and Large Language Models training~\cite{zhao2023survey}.

Depending on whether they find all valid set pairs, R-S Join algorithms are categorized as {\em approximate} or {\em exact} computation. 
Approximate computation exploits well-designed set similarity estimation functions to accelerate computing, e.g., edit distance based~\cite{Yang2022ICDE,Karpov2024}, signature-scheme based~\cite{Schmitt2023}, high-dimensional data sketching and indexing based~\cite{christiani2018scalable}. Such algorithms may not output all satisfied set pairs since they sacrifice accuracy for efficiency. 

Exact R-S Joins are important in {\em large-scale metagenomics analysis} (i.e., identifying near-duplicates in massive genome samples)~\cite{besta2020communication} 
, money laundering~\cite{tian2025towards},
and financial fraud detection~\cite{metwally2024similarity}.
Such scenarios call for exact computation, which outputs all satisfied set pairs, e.g., {\em All-Pairs}~\cite{bayardo2007scaling}, {\em PPJoin}~\cite{xiao2011efficient}, and {\em MetricJoin}~\cite{widmoser2023metricjoin}. Such algorithms utilize the {\em filter-and-verification} framework to accelerate the computation, which consists of 2 stages. 
The \(1^{st}\) stage is the {\em filtering stage}, which 
applies appropriate filters to prune candidates, 
e.g., {\em the length filter} in {\em All-Pairs}~\cite{bayardo2007scaling}, position and suffix filters in {\em PPJoin} and {\em PPJoin+}~\cite{xiao2011efficient}, the prefix filter in AdaptJoin~\cite{wang2012can}, and {\em the bitmap filter}~\cite{sandes2020bitmap}. The \(2^{nd}\) stage, i.e., {\em verification stage}, computes the similarity of the candidates and outputs satisfied pairs. R-S Joins with filter-and-verification have demonstrated their remarkable performance.
 
The large scale of modern datasets, however, raises great challenges to sequential R-S Join algorithms that are limited
by the number of threads supported by the processing units. 
To handle these data volumes, distributed R-S Join algorithms have been proposed~\cite{rong2017fast,fier2018set,metwally2024similarity}.
They build on the filter-and-verification paradigm and leverage MapReduce~\cite{dean2008mapreduce} to split large datasets into smaller splits that are dispatched to cluster nodes for parallel processing.

\begin{figure}[tb]
\centering
{\includegraphics[width=0.45\textwidth]{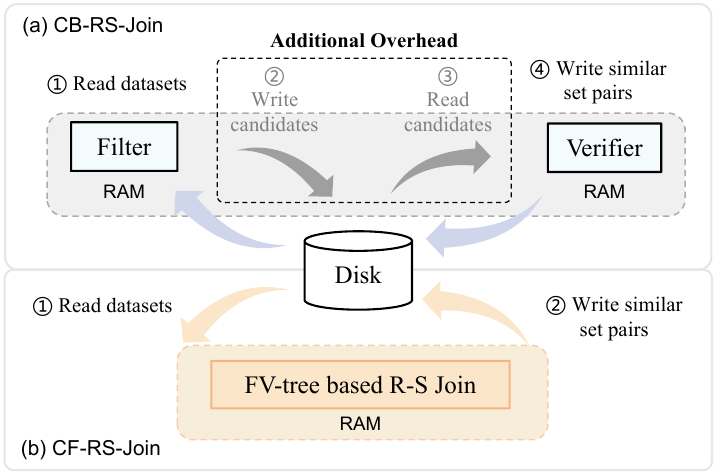}}
\caption{Candidate-Based vs. Candidate-Free R-S Joins}
\label{fig:motivation}
\end{figure} 
 
As depicted in Fig.~\hyperref[fig:motivation]{\ref*{fig:motivation}(a)}, in the existing distributed R-S Join with filter-and-verification framework, the filters and verifiers are independent execution units. This process involves four major I/O stages: filters read input data from disks (\ding{172}), write the generated candidates to disks (\ding{173}), then verifiers read these candidates from disks (\ding{174}), and finally write the satisfied pairs to disks as output (\ding{175}). Such approaches are called candidate-based R-S Join ({\em CB-RS-Join}). Excessive candidates generated by CB-RS-Join for large-scale set collections lead to high I/O and pairwise verification costs, which degrade the overall performance.

Our approach is to devise filter-and-verification trees ({\em FV-trees}, or {\em FVTs}) to compress one collection of sets in a compact data structure and keep it in memory, then scan the data structure using another collection for R-S Join computation. 
By integrating the filtering and verification into one stage, as shown in Fig.~\hyperref[fig:motivation]{\ref*{fig:motivation}(b)}, our approach reduces the I/O operations to only two: reading the input datasets (\ding{172}) and writing the final results (\ding{173}).
This removal of the candidate generation phase is why we call our approach candidate-free R-S Join ({\em CF-RS-Join}), and it targets a significant acceleration of the computation. To the best of our knowledge, this work is the first candidate-free exact R-S Join.

Our main contributions can be summarized as follows:
\begin{itemize}
    \item Propose an FVT recording of elements and their associated sets that compresses data in memory, enabling fast lookups and reducing database scan frequency. Then, an original candidate-free R-S Join based on the FVT, named CF-RS-Join/FVT, is proposed, and its theoretical proof of correctness is elaborated. CF-RS-Join/FVT integrates the filtering and verification in one single stage, eliminating the candidate generation.
    \item Design a more compact FVT variant, the Linear FVT (LFVT), which reorganizes nodes along non-branching paths in the FVT into a single node, and corresponding data are compressed in a linear array. LFVT based R-S Join algorithm, i.e., CF-RS-Join/LFVT, has been proposed to improve the traversal efficiency further. 
    \item Design {\em MR-CF-RS-Join/FVT} and {\em MR-CF-RS-Join/LFVT} to extend our approaches using MapReduce for parallel and distributed processing, further improving performance.
    \item Conduct empirical studies on 7 real-world datasets to evaluate the proposed algorithms against selected state-of-the-art (SOTA) algorithms, and results show that the MR-CF-RS-Join/LFVT achieves the best performance.
\end{itemize} 

The rest of the paper is organized as follows: Section~\ref{sec:related work} investigates related work on exact R-S Joins, Section~\ref{sec:A-tree based method} introduces the design of the FVT, LFVT, and their corresponding candidate-free R-S Join algorithms CF-RS-Join/FVT and CF-RS-Join/LFVT. Section~\ref{sec:FastAP-treeR-SJ} describes MapReduce and FVT/LFVT based R-S Join algorithms MR-CF-RS-Join/FVT and MR-CF-RS-Join/LFVT. 
Section~\ref{sec:Experimental Evaluation} reports the empirical comparison study, 
and Section~\ref{sec:conclusion} concludes the paper. 

\section{Related Work}
\label{sec:related work}
R-S Join has been extensively studied for about 20 years and remains an active research topic. Exact R-S Joins require measuring the similarity between all pairs of sets in two collections, with the time complexity $O(k \times m \times n)$, where $m$ and $n$ denote the size of two collections, respectively, and $k$ is the average computation time for the similarity between any two sets. Large-scale collections pose challenges for R-S Joins with such high computational complexity. 
To address this challenge, 
existing works employing the filter-and-verification framework leverage in-memory data representation and parallel/distributed computing to accelerate the exact R-S Join computation.

According to the techniques used to represent data in memory to expedite the computation, existing R-S Joins can be classified into {\em inverted index based} and {\em tree based}. 
When an inverted index based algorithm is applied, different filter strategies are first devised to generate prefixes~\cite{xiao2011efficient,bouros2012spatio,wang2012can,wang2017leveraging} or signatures (partitioned disjoint subsets of a set)~\cite{deng2015efficient} to construct an {\em inverted index}. This index then prunes dissimilar set pairs and obtain candidates, whose real similarity is finally computed in the verification stage.
Recently, a lossless string similarity compression technology {\em CSS} (compressed string similarity) has been proposed to reduce the inverted index size, reducing the memory consumption by 3 to 5 times for efficient R-S Join computation~\cite{Xiao2022ICDE}. 
Other R-S Join algorithms that focus on approximate~\cite{broder1997resemblance,christiani2018scalable,indyk1998approximate,satuluri2011bayesian} and parallel~\cite{fier2022parallelizing,ribeiro2017fast} are beyond the scope of this paper.
When a tree-based algorithm is applied, e.g., tree structures are applied to represent data in memory for set pair lookups, e.g., B$^{+}$-tree like structure~\cite{zhang2017efficient}, 
and R-trees~\cite{widmoser2023metricjoin}.
Such algorithms first organize data into a tree index, and each set is probed against the tree index to find similar set pairs. 
R-tree based MetricJoin maps sets from a general metric space to a multiple-dimensional vector space to build R-trees, and exploits metric properties of the set distance to prune unqualified sets and reduce the number of candidates. 

The parallel nature of modern multi-core computers makes multi-threading an important technique for accelerating R-S Join, harnessing the multi-core power of CPU like multi-threading PPJoin and AllPairs~\cite{fier2022parallelizing}, and many cores of Graphics Processing Units (GPUs) like gSSJoin \cite{junior2016gssjoin} and fgssjoin \cite{quirino2018efficient}. 
Meanwhile, CPU-based cluster computing is a more inexpensive and prevalent form of parallel distributed computing. 
Most existing parallel distributed set similarity join algorithms for CPUs rely on the MapReduce framework for execution acceleration. 
A MapReduce cluster consists of multiple computers with a shared-nothing architecture, and a MapReduce job includes 2-stage computation: {\em map stage} and {\em reduce stage}. The input data is sliced into multiple splits, each of which will be sent to a computer in the cluster, i.e., {\em mapper}, for executing \texttt{map()} function in parallel. The intermediate results of \texttt{map()} functions will be combined and shuffled to one or multiple computers in the cluster, i.e., {\em reducer}, for \texttt{reduce()} function execution. 
An R-S Join usually includes two or more MapReduce jobs. According to whether a set is sliced into segments for parallel processing, existing distributed R-S Join algorithms can be classified into two categories: (1) {\em Entire set filter-and-verification based algorithms}: The entire set is dispatched to a node for filtering or verification, e.g., RIDPairsPPJoin~\cite{vernica2010efficient}, MGJoin~\cite{rong2012efficient}, SSJ-2R~\cite{baraglia2010document}, and FSSJ~\cite{metwally2024similarity}; and (2) {\em Set segment filter-and-verification based algorithms}: A set is sliced into segments, each of which will be dispatched to a node for filtering, then intermediate results will be merged for verification, e.g., FS-Join~\cite{rong2017fast}. 

Both categories of existing distributed R-S Joins belong to Candidate-Based R-S Joins (CB-RS-Joins).
However, when processing large-scale set collections, this approach generates an excessive number of candidates. This leads to high costs in disk I/O, data transmission, and pairwise verification, which ultimately degrades overall performance.  

\section{Candidate-Free R-S Joins (CF-RS-Joins)} \label{sec:A-tree based method}

To eliminate the candidate generation for better R-S Join computation efficiency, our solution is to design tree structures to compress elements and their associated set into memory and integrate filtering and verification into one single stage.

\begin{figure*}[th]
\centering
{\includegraphics[width=1\textwidth]{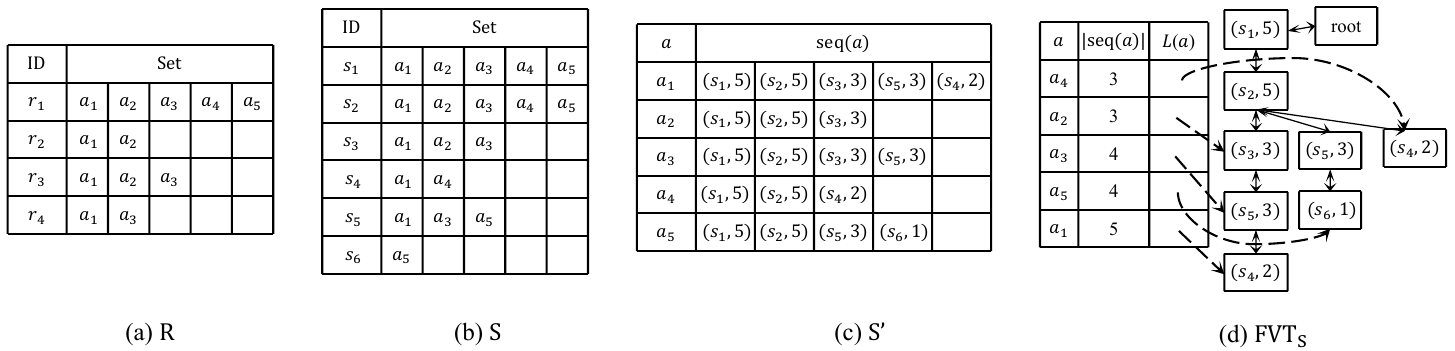}}
\caption{Set collections \(\text{R}\), \(\text{S}\), \(\text{S}'\) (reorganized from $\text{S}$), and the FVT \(\texttt{FVT}_{\text{S}}\) constructed over $\text{S}'$}
\label{fig:sample_collections}
\end{figure*} 
Let the universe of elements be \(\mathcal{A}\), 
given two collections of sets $\mathcal{R}$ and $\mathcal{S}$, for any \(r \in \mathcal{R}\) and \(s \in \mathcal{S}\), we have \(r \subseteq \mathcal{A}\) and \(s \subseteq \mathcal{A}\). 
For any \( r \in \mathcal{R}\) and \(s \in \mathcal{S}\), their similarity, denoted as \(\texttt{sim}(r, s)\), can be measured using similarity coefficients like Jaccard, Overlap, Cosine, and Dice. All require \(|r|\), \(|s|\), and \(|r \cap s|\).
Table~\ref{tab:summary of notations} summarizes the notations, and Table~\ref{tab:sim coefficients} lists the formulas of the similarity coefficients.
To put the discussion into perspective, this paper uses Jaccard to demonstrate how our proposed algorithm works, the same computation can be applied when the other three coefficients are used.

\begin{table}[tb]
    \centering
     \caption{Summary of notations}
     \resizebox{\linewidth}{!}{
        \begin{tabular}{cc}
        \toprule
            Notation  & Definition \\
        \midrule
                        $ t $ & similarity threshold \\
            \(\text{seq}(a)\) & sequence of all sets containing element \(a\) \\
            $\texttt{FVT}$ & filter-and-verification tree structure \\
            $\mathcal{E}$ & element table mapping elements to tree nodes \\
            $ \mathcal{T} $ & tree representation of FVT/LFVT \\
            $ par $ & parent node of a tree node \\
            $ cld$ & set of child nodes of a tree node \\
            $ n $ & node in FVT/LFVT \\
            $ n_{s} $ & tree node representing set $s$ in $\texttt{FVT}$ \\
            \(path(n_u, n_v)\) & path in FVT/LFVT from node \(n_u\) to node \(n_v\) \\
            $ \mathcal{P} $ & set of similar pairs meeting the threshold \\
            $ \mathcal{F} $ & map from a set $s$ to its intersection size with a set $r$, i.e., $f_{r,s}$ \\
            $ f_{r,s} $ & intersection size of set pair $(r, s)$ \\
            $k$ & number of reducers \\
            $\lambda_b$ & beginning (lower) bound of a set-length partition \\
            $\lambda_e$ & ending (upper) bound of a set-length partition \\
            $\text{load}(\cdot)$ & workload estimation function for a reducer \\
            $\psi(\cdot)$ & optimal cost (minimized maximum load) for a data partitioning strategy \\
        \bottomrule
        \end{tabular}
     }
\label{tab:summary of notations}
\end{table}

\begin{table}[tb]
\centering
 \caption{
 The computation of similarity coefficients }
    \begin{tabular}{cccc}
    \toprule
       Coefficients  &  \(\texttt{sim}(r,s)\) & $lb_r$ & $ub_r $ \\
    \midrule
        Jaccard & \(
        \frac{|r \cap s|}{|r \cup s|} 
        = \tfrac{|r \cap s|}{|r| + |s| - |r \cap s|}
        \) & $t \times |r|$ &  $\frac{|r|}{t}$\\
        Overlap & \(|r \cap s|\) &  $t$  & $+\infty$   \\
  Cosine& \( \frac{|r \cap s|}{\sqrt{|r| \times |s|}}\)&  $t^2 \times |r|$   & $\frac{|r|}{t^2}$     \\
        Dice & \(\frac{2\ \times |r \cap s|}{|r| + |s|}\) &  $\frac{t \times |r|}{2-t}$ & $\frac{(2-t)\times |r|}{t}$  \\
    \bottomrule
    \end{tabular}
    \label{tab:sim coefficients}
\end{table}

For clarity, we formally define  the R-S set similarity join: 
\begin{definition}[R-S Set Similarity Join]
Given two set collections $\mathcal{R}$ and $\mathcal{S}$, a similarity threshold $t \in (0,1]$,
and similarity function $\texttt{sim}(r,s)$,
the R-S set similarity join identifies all pairs $(r,s) \in \mathcal{R} \times \mathcal{S}$ 
that satisfy $\texttt{sim}(r,s) \geq t$.
\end{definition}

\begin{example}
Assume $\text{R}=\{r_1=\{1,2,3\},r_2=\{3,4\}\}$ and $\text{S}=\{s_1=\{1,3,4\},s_2=\{2,3\},s_3=\{1,2\}\}$ with a Jaccard similarity threshold $t=0.6$. The objective is to find all pairs $(r,s)$ where $\texttt{sim}(r,s) \ge 0.6$. By calculating the Jaccard similarity for all pairs, we find that $\texttt{sim}(r_1,s_2)=\texttt{sim}(r_1,s_3)=\texttt{sim}(r_2,s_1)=2/3$, all of them meet the threshold. Therefore, the final result is $\{(r_1,s_2),(r_1,s_3),(r_2,s_1)\}$.
\end{example}

As a running example, we use two sample collections $\text{R}$ (Fig.~\hyperref[fig:sample_collections]{\ref*{fig:sample_collections}(a)}) and $\text{S}$ (Fig.~\hyperref[fig:sample_collections]{\ref*{fig:sample_collections}(b)}) to illustrate our candidate-free R-S Joins (CF-RS-Joins).
\subsection{CF-RS-Join with Filter-and-Verification Tree}
\label{sucsec:tree based structure}

This section describes how filtering and verification are integrated into a single stage for R-S Join using the CF-RS-Join with filter-and-verification tree (FVT) (CF-RS-Join/FVT). We first introduce the FVT data structure and its construction process, and then present the R-S Join computation using the constructed FVT.
Finally, we provide a comprehensive complexity analysis.

\subsubsection{FVT Structure and Construction}
\label{subsubsec:AP-tree structure}
An FVT compresses elements and their associated sets in memory for R-S Join computation. 
The FVT constructed from collection $\mathcal{S}$ is denoted as \(\texttt{FVT}_{\mathcal{S}} = (\mathcal{E}_{\mathcal{S}}, \mathcal{T}_{\mathcal{S}})\), where \(\mathcal{E}_{\mathcal{S}}\) is an element table that enables fast lookup of sets having a particular element in $\mathcal{R}$, and \(\mathcal{T}_{\mathcal{S}}\) is a tree whose node $n_s$ represents a set \(s \in \mathcal{S}\). Each node $n_s$ is a 3-tuple, i.e., $n_s=((s,|s|),cld,par)$, where $cld$ is the set of child pointers in \(\mathcal{T}_{\mathcal{S}}\) and $par$ is the parent pointer. 
In particular, \(cld=\emptyset\) means that the node has no child, and \(par=\text{NULL}\) means that the node is the root node. 
Meanwhile, the root node is denoted as $n_\text{root}=((\emptyset,0),cld,\text{NULL})$. 
In addition, for any node $n_{s}$ on a path from the root to a leaf, the larger the \(|s|\), the closer the node \(n_{s}\) is to the root. This is to guarantee nodes with bigger sets are more likely to be shared in the tree, which compresses data representation and makes the tree compact.

As illustrated in Fig.~\ref{fig:sample_collections}, the construction of an FVT proceeds in two steps with a single database scan.

\noindent{\bf Step 1: Reorganize the set collection.} We transform $\mathcal{S}$ into a mapping from each element $a$ to an ordered sequence $\text{seq}(a)$.
This sequence consists of 2-tuples $(s,|s|)$ for all $s \in \mathcal{S}$ where $a \in s$,
sorted by non-increasing $|s|$ so that larger sets appear first.
The resulting reorganized collection, denoted $\mathcal{S}'$, is the structure over which the FVT is constructed in Step 2.

\begin{example}
For an element \(a_5\) in Fig.~\hyperref[fig:sample_collections]{\ref*{fig:sample_collections}(c)}, it is in sets \(s_1\), \(s_2\), \(s_5\), and \(s_6\), then the value of key \(a_5\), i.e., \(\text{seq}(a_5)\), is an ordered sequence \(\text{seq}(a_5) = \langle (s_1,5),(s_2,5),(s_5,3),(s_6,1)\rangle\).
\end{example}

\noindent{\bf Step 2: Construct the FVT over the reorganized data.}
The FVT constructed over the reorganized data \(\mathcal{S}'\) is denoted as \(\texttt{FVT}_{\mathcal{S}}\). 
This process involves building a tree structure \(\mathcal{T}_{\mathcal{S}}\) and an element table \(\mathcal{E}_{\mathcal{S}}\). 
For each entry \((a, \text{seq}(a))\) in \(\mathcal{S}'\), let the \(\text{seq}(a) = \langle (s_{a}^1, |s_{a}^1|), \ldots, (s_{a}^{|\text{seq}(a)|}, |s_{a}^{|\text{seq}(a)|}|) \rangle\), where \(s_{a}^i\) represents $i$-th tuple in $\text{seq}(a)$ ($1 \le i \le |\text{seq}(a)|$). We perform two operations:
\begin{enumerate}
    \item {\bf Path creation.} 
    For the sequence $\text{seq}(a)$, we ensure a corresponding path exists in \(\mathcal{T}_{\mathcal{S}}\).
    Starting from the $n_{\text{root}}$,
    we traverse or create nodes for each tuple \((s_a^i, |s_a^i|)\) in $\text{seq}(a)$, in order.
    If a prefix of the path already exists, we traverse it; otherwise, new nodes are appended to the longest existing prefix.
    The resulting path, from the root $n_{\text{root}}$ to the terminal node $n_{s_a^{|\text{seq}(a)|}}$, is denoted as $path(n_{\text{root}}, n_{s_{a}^{|\text{seq}(a)|}}) = \langle n_{\text{root}}, n_{s_a^1}, \ldots, n_{s_a^{|\text{seq}(a)|}} \rangle$.
    \item {\bf Element table insertion.} We add a 3-tuple \((a, \allowbreak |\text{seq}(a)|, \allowbreak L(a))\) to the element table \(\mathcal{E}_{\mathcal{S}}\), where \(L(a)\) is a pointer to the terminal node \(n_{s_a^{|\text{seq}(a)|}}\) of this path created for element $a$.
\end{enumerate}

\begin{example}
For the first entry in \(\text{S}'\) (Fig.~\hyperref[fig:sample_collections]{\ref*{fig:sample_collections}(c)}), $\text{seq}(a_1)  = \langle (s_1,5), (s_2,5), (s_3,3), (s_5,3), (s_4,2)\rangle$, 
a new path $path(n_\text{root}, n_{s_4})=\langle n_\text{root}, n_{s_1}, n_{s_2}, n_{s_3}, n_{s_5}, n_{s_4}\rangle$ is added to \(\mathcal{T}_{\text{S}}\). 
For the third entry in \(\text{S}'\), $\text{seq}(a_3) = \langle (s_1,5), (s_2,5), (s_3,3), (s_5,3) \rangle$, the path corresponding to this sequence, $path(n_\text{root}, n_{s_5})$, is a prefix of the path created for $a_1$. Thus, no new node is added to \(\mathcal{T}_{\text{S}}\). 
Only a new 3-tuple is added to \(\mathcal{E}_{\text{S}}\): \((a_3, 4, L(a_3))\), where \(L(a_3)\) points to \(n_{s_5}\).
\end{example} 

\subsubsection{The CF-RS-Join/FVT Algorithm}
\label{subsubsec:AP-treeR-SJ}
This section presents CF-RS-Join/FVT, an optimized R-S set similarity join algorithm. 
Existing methods employ separate filtering and verification stages, inevitably generating excessive intermediate candidates that incur substantial computational and I/O overhead. 
CF-RS-Join/FVT overcomes it by first constructing an FVT for one collection and then 
traversing this tree to perform the join using another collection, unifying the filtering and verification stages. 
The traversal incorporates a length filter strategy for early pruning of dissimilar set pairs. 
We begin by detailing this length filter strategy, followed by a formal proof of the algorithm’s correctness.

CF-RS-Join/FVT utilizes the length filter to prune dissimilar set pairs, as formalized in Lemma \ref{lma:len filter}. 
The corresponding proof is provided in \cite{rong2017fast}.
This strategy, based on the observation that sets with significantly different sizes cannot be similar, 
is widely adopted in R-S Join methods~\cite{widmoser2023metricjoin}. 

\begin{lemma}[Length Filter]
\label{lma:len filter}
    Given two sets $r$ and $s$, and a threshold $t$, if $\texttt{sim}(r, s) \ge t$, then 
    $ lb_r  \le |s| \le  ub_r $. When $\texttt{sim} = \text{Jaccard}$, $lb_r =\lceil |r|\times t\rceil, ub_r= \lfloor |r|/t \rfloor$ (cf. Table \ref{tab:sim coefficients}).
\end{lemma}

The core computation of CF-RS-Join/FVT is the efficient calculation of set intersection sizes $|r \cap s|$ via FVT traversal, which we measure using a counter $f_{r,s}$.
The theoretical foundation for this computation is formalized in Lemma \ref{lemma2}.
For clarity, we temporarily omit the length filter to focus on the fundamental intersection calculation.

\begin{lemma}[Correctness of CF-RS-Join/FVT w/o Length Filter]
\label{lemma2}
Given a set $r\in \mathcal{R}$, for each element $a \in r$ present in the element table $\mathcal{E}_{\mathcal{S}}$, 
we traverse the tree $\mathcal{T}_{\mathcal{S}}$ from the node pointed to by $L(a)$ up to the root. During the traversal,
every visited node corresponds to a specific set $s_i$ that contains $a$. For each such node, we increment the corresponding counter: $f_{r,s_i} \gets f_{r,s_i} + 1$.
After processing all elements in $r$, the final value of any counter $f_{r,s}$ equals the intersection size $|r \cap s|$.
\end{lemma}

\begin{proof}
For any \(r \in \mathcal{R}\) and any $s \in \mathcal{S}$, $f_{r,s}$ is initialized to be 0. For any \(r \in \mathcal{R}\), and for any $a \in r$:
\begin{itemize}
    \item \textbf{Case 1:} If $(a,|\text{seq}(a)|, L(a)) \notin \mathcal{E}_{\mathcal{S}}$, 
    then $\forall s\in \mathcal{S}$, $a \notin s$ and $a \notin r \cap s$.
    \item \textbf{Case 2:} If $(a,|\text{seq}(a)|, L(a)) \in \mathcal{E}_{\mathcal{S}}$, 
    traverse \(\mathcal{T}_{\mathcal{S}}\) from $L(a)$ to $n_\text{root}$. 
    For each visited node corresponding to $(s,|s|) \in \text{seq}(a)$, 
    since $a \in s$ and $a \in r$, we have $a \in r \cap s$. 
    Thus, increment $f_{r,s}$.
\end{itemize}
Based on the above discussion, during the traversal from the node pointed to by $L(a)$ to the root node $n_\text{root}$ along each node's parent pointer, the operation $f_{r,s} \leftarrow f_{r,s} + 1$ is performed
if and only if
$a \in r \cap s$. Since $a$ is unique in the element table \(\mathcal{E}_{\mathcal{S}}\), after the traversal, $f_{r,s}$ equals $|r \cap s|$, the size of the intersection.
\end{proof}

We observe that during the traversal process in Lemma \ref{lemma2}, for every path from the node pointed to by $L(a)$ to the root node $n_\text{root}$, the set size $|s|$ in each node follows a non-decreasing order. Combining this structural property with Lemma \ref{lma:len filter}, we establish Theorem \ref{theorem}.

\begin{theorem}[Correctness of CF-RS-Join/FVT w/ Length Filter]
\label{theorem}
For any \(r \in \mathcal{R}\) and any $a \in r$, consider the path from the node pointed to by $L(a)$ up to the root, denoted as \(path(L(a), n_\text{root})\).  
If \(n_{s_x}\) is the first node in \(path(L(a),n_\text{root}) = \langle L(a),L(a).par, \ldots, n_{s_x},\ldots, n_\text{root}\rangle\) with a set size $|s_x| > ub_r$, the traversal terminates early, because any node closer to the root will also have a set size greater than $ub_r$.
After the traversal is completed, we have all set pairs with similarity no less than the threshold.
\end{theorem}

\begin{proof}
The correctness of the algorithm without early stopping is established by Lemma~\ref{lemma2}. We need to show that the early-stop optimization does not discard any valid result pairs.
The proof relies on a key structural property of the FVT: along any path from a node to the root, the set sizes are monotonically non-decreasing.
The algorithm applies the length filter in two ways: it ignores nodes where $|s| < lb_r$ and continues traversal, but it terminates the traversal path for an element $a$ when it first encounters a node $n_{s_x}$ where $|s_x| > ub_r$. 
If the traversal stops at $n_{s_x}$ because $|s_x| > ub_r$, then by Lemma~\ref{lma:len filter}, the pair $(r, s_x)$ cannot satisfy the similarity threshold. 
Now, consider any subsequent node $n_{s_y}$ on the path towards the root. Due to the FVT's monotonic property, $|s_y| \ge |s_x|$, 
it necessarily follows that $|s_y| > ub_r$. Thus, $(r, s_y)$ and all subsequent pairs on this path will also fail the length filter condition. 
Therefore, terminating the traversal once the upper bound is exceeded is a safe optimization.
\end{proof}

The core join process of CF-RS-Join/FVT, whose correctness is established by Lemma~\ref{lemma2} and Theorem~\ref{theorem}, is as follows: for each set $r \in \mathcal{R}$ and for each element $a \in r$, the algorithm traverses the FVT, starting from the node indexed by $L(a)$ and moving upwards towards $n_\text{root}$. During this traversal, the length filter is applied at each node $n_s$ to prune the search space. 
\begin{itemize}
    \item \(|s| < lb_r\): Skip $n_s$ and proceed to its parent; \(f_{r,s}\) remains unchanged.
    \item \(lb_r \le |s| \le ub_r\): Increment \(f_{r,s}\). If the parent of \(n_s\) is not the root, set \(n_s\) to its parent and continue; otherwise, end the traversal for \(a\).
    \item \(|s| > ub_r\): Terminate the current traversal and proceed to the next element in \(r\), since any ancestor will have an even larger set size in the FVT. 
\end{itemize}
This process is repeated for all elements in $r$ to compute the final intersection sizes.

\begin{example}
\label{exm:CF-RS-Join/FVT}
We illustrate the query processing using $r_4 = \{a_1, a_3\}$ and $t=0.6$ on the FVT from Fig.~\hyperref[fig:sample_collections]{\ref*{fig:sample_collections}(d)}.  
The length filter restricts candidates in $\text{S}$ to the range $[2,3]$.
The algorithm iterates over $r_4$ to populate the frequency map $\mathcal{F}$.
For $a_1$, the traversal from $L(a_1)$ visits valid nodes $n_{s_4}, n_{s_5}, n_{s_3}$ (all within the length range), the $f_{r_4,s}$ increments are as follows: $f_{r_4,s_4}=1$, $f_{r_4,s_5}=1$, and $f_{r_4,s_3}=1$. 
For $a_3$, the traversal updates counts for $s_5$ and $s_3$, resulting in $f_{r_4,s_5}=2$ and $f_{r_4,s_3}=2$.
The algorithm terminates the path upon reaching a node like $n_{s_2}$ whose set size (5) is outside the valid range.
After processing all elements in $r_4$, the final intersection counts in $\mathcal{F}$ are: $f_{r_4,s_4}=1$, $f_{r_4,s_5}=2$, and $f_{r_4,s_3}=2$. This result now correctly reflects the true intersection sizes: $|r_4 \cap s_4|=1$, $|r_4 \cap s_5|=2$, and $|r_4 \cap s_3|=2$.
\end{example}

To implement this join process efficiently, the CF-RS-Join/FVT algorithm employs a strategy to process shared paths effectively and minimize redundant computations. This is crucial when multiple elements from a set $r$ share common ancestral nodes in the FVT. The complete procedure is detailed in Algorithm~\ref{alg2}.

\begin{algorithm}[tb]
    \footnotesize
    \caption{The CF-RS-Join/FVT algorithm}
    \label{alg2}
    \KwIn{Two set collections $\mathcal{R}$ and $\mathcal{S}$, a threshold $t$}
    \KwOut{\small $\mathcal{P} = \{ (r,s) \mid r \in \mathcal{R}, s \in \mathcal{S}, \operatorname{Jaccard}(r, s) \geq t\}$}
    Construct $\texttt{FVT}_{\mathcal{S}} = (\mathcal{E}_{\mathcal{S}}, \mathcal{T}_{\mathcal{S}})$ over $\mathcal{S};$
    
    Initialize $\mathcal{P}\gets \emptyset;$
    
    \For{\textnormal{each} $r \in \mathcal{R}$}{
        $\mathcal{F}\gets$ \{\};\tcp{Map: $s \to f_{r,s}$}
        $\mathcal{N}\gets$ \{\};\tcp{Map: $L(a) \to$ \#occurrences in $\mathcal{T}_{\mathcal{S}}$}
        $lb_r \gets \lceil |r|\times t\rceil$; $ub_r \gets \lfloor |r|/t \rfloor;$

        \For{ \textnormal{each} $a \in r$}{
            \If{$a \in \mathcal{E}_{\mathcal{S}}$}{
                $node \gets L(a)$ \textnormal{from} $\mathcal{E}_{\mathcal{S}};$
                
                $\mathcal{N}[node] \gets \mathcal{N}[node] + 1;$
            }
        }
        
        Sort elements in $r$ by $\mathcal{E}_{\mathcal{S}}[a].|\text{seq}(a)|$ in descending order;

        \For{ \textnormal{each} $a \in r$}{
            \If{$\mathcal{N} = \emptyset$}{
                break\;
            }

            \If{$L(a) \notin \mathcal{N}$}{
                        continue\;
                    }
        
            $node \gets L(a);$
            
            $support \gets \mathcal{N}[node]$;
            
            Remove $node$ from $\mathcal{N}$;
            
            \While{ $node \neq n_\textnormal{root}$ \textnormal{and} $node.|s| \leq ub_r$} {
                \If{$node \in \mathcal{N}$}{
                    
                    $support \gets support+\mathcal{N}[node];$
                    
                    Remove $node$ from $\mathcal{N};$
                }
                \If{$node.|s| \ge lb_r$}{
                    
                    $s \gets \text{get\_set}(node);$
                    
                    \If{$s \notin \mathcal{F}$}{
                        $\mathcal{F}[s] \gets 0;$
                    }
                    $\mathcal{F}[s] \gets \mathcal{F}[s] + support;$
                }
                $node \gets node.par;$
            }
        }
        \For{\textnormal{each} $(s, f_{r,s}) \in \mathcal{F}$}{
            Calculate $\operatorname{Jaccard}(r,s)$ using $f_{r,s};$
            
            \If{$\operatorname{Jaccard}(r,s)\ge t$}{
                $\mathcal{P} \gets \mathcal{P} \cup \{(r,s)\};$
            }
        }
    }
\end{algorithm}

The operational flow of the algorithm is detailed as follows. 
For a given query set $r$, 
the algorithm first retrieves the corresponding FVT nodes $L(a)$ for all $a \in r$ and records their occurrence counts in a map $\mathcal{N}$ (lines 7-12).
This map serves to detect whether an ancestor node during traversal is also a starting node for other elements in $r$.
Then, the elements in $r$ are sorted by descending order of $|\text{seq}(a)|$ (line 13), which ensures that elements corresponding to deeper nodes in the FVT are processed first.
The main loop processes elements of $r$ in the sorted order (lines 14-38). 
For each element $a$, if its starting node $L(a)$ exists in $\mathcal{N}$, a support counter is initialized with its frequency from $\mathcal{N}$ (lines 21-23).
As the algorithm traverses upwards to the root, it dynamically aggregates support from other elements of $r$ that lie on the current path (lines 25-28).
Specifically, if an ancestor node corresponds to another element in $r$ (i.e., exists in $\mathcal{N}$), 
its frequency is added to the current support, and the node in $\mathcal{N}$ is cleared to prevent redundant processing. 
This mechanism ensures that shared path segments are traversed only once, with the support value correctly representing the collective contribution of all relevant elements in $r$ (line 34). 
The length filter is applied at each step to facilitate early termination (lines 24, 29). 
After all elements in $r$ are processed, the similarities are verified for pairs with non-zero intersection sizes (lines 39-44).

Based on the aforementioned discussion, the benefits of CF-RS-Join/FVT,  
can be summarized as: (1) The element table facilitates fast element-set lookups; 
(2) It integrates the two-stage filter-and-verification into a single traversal stage, using the filter for early termination (as set sizes increase toward the root);
(3) It eliminates redundant database scans by maintaining the FVT in memory.

\textbf{Time complexity analysis.}
The computation complexity of CF-RS-Join/FVT comprises two phases: FVT construction and R-S Join computation.
The FVT construction phase involves reorganizing the set collection \(\mathcal{S}\) into an inverted index \(\mathcal{S}'\), which takes $O(\sum_{s \in \mathcal{S}} |s|) = O(|\mathcal{S}| \times \overline{|s|})$ time. Then, the FVT is built by inserting each of the \(|\mathcal{S}'|\) element sequences. This step's time is proportional to the sum of all sequence lengths, i.e., $O(\sum_{a \in \mathcal{S}'} |\text{seq}(a)|) = O(|\mathcal{S}'| \times \overline{|\text{seq}(a)|})$.
The R-S join computation phase consists of two main steps for each set \(r \in \mathcal{R}\): FVT traversal and similarity verification. The cost of the verification step is determined by the number of candidates collected, which is bounded by the total number of nodes visited during the traversal. Consequently, the overall cost for this phase is dominated by the traversal, which for a set \(r\) is bounded by $\sum_{a \in r} |\text{seq}(a)|$. Thus, the total join computation cost is $O(\sum_{r \in \mathcal{R}} \sum_{a \in r} |\text{seq}(a)|) = O(|\mathcal{R}| \times \overline{|r|} \times \overline{|\text{seq}(a)|})$.
In all, the overall time complexity is $O(|\mathcal{S}| \times \overline{|s|} + |\mathcal{S}'| \times \overline{|\text{seq}(a)|} + |\mathcal{R}| \times \overline{|r|} \times \overline{|\text{seq}(a)|})$.
In practice, one may build the FVT on the collection that minimizes the expected traversal cost (e.g., the sum of $|\text{seq}(a)|$). 

\begin{figure*}[th]
\centering
{\includegraphics[width=1\textwidth]{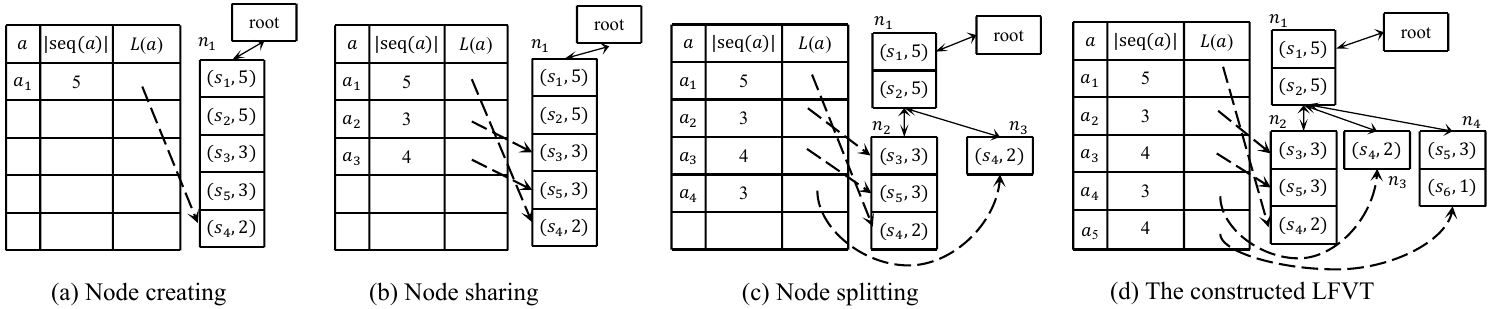}}
\caption{The construction of an LFVT over \(\text{S}\)}
\label{fig:telpres-tree build}
\end{figure*} 
\subsection{CF-RS-Join with Linear FVT}
\label{subsec:LAP-tree structure} 

Although FVT enables efficient candidate-free R-S Join computation, its tree structure suffers from poor cache locality during traversal due to the non-contiguous memory layout of nodes along a path.
We observe that many paths in FVT contain no branching, e.g., \(path(n_{s_2},n_{s_1})\) in Fig.~\hyperref[fig:sample_collections]{\ref*{fig:sample_collections}(d)}. 
Merging such paths into single nodes can improve both cache locality and memory efficiency. 
This section introduces Linear FVT (LFVT), a more compact FVT variant that compresses non-branching paths, and the corresponding CF-RS-Join/LFVT algorithm.

An LFVT contains an element table $\mathcal{E}$ and a tree $\mathcal{T}$ , where the design philosophy of the element table is exactly the same as the FVT. 
There are differences in the structure of the tree node and the actual pointing of the \(L(a)\). 
First, a tree node, denoted by an indexed symbol such as $n$, is represented as a 3-tuple $n = (T, cld, par)$. 
Here, $T$ is the ordered sequence of 2-tuples in \(\mathcal{S}'\). 
Let $T = \langle (s^1, |s^1|), (s^2, |s^2|), \ldots, (s^m, |s^m|) \rangle$, we have \(|s^{x+1}| \leq |s^{x}|\) for $1 \le x < m$. 
The root node always has an empty sequence, i.e., $n_\text{root}=(\emptyset,cld,NULL)$. 
Similar to FVT, except for the root node, 
sets in a node's sequence have larger sizes than those in its child node (if any).
Second, the $L(a)$ in a 3-tuple of the element table points to a particular 2-tuple of a node's sequence in $n$, while in an FVT $L(a)$ points to a node.

Fig. \ref{fig:telpres-tree build} illustrates LFVT construction for the reorganized collection of tuples in \(\text{S}'\) depicted in Fig. \ref{fig:sample_collections}. 
First, the LFVT is initialized with a root node and an empty element table.  
The first sequence, $\text{seq}(a_1)$, creates a new node $n_1$ to store it, as shown in Fig.~\hyperref[fig:telpres-tree build]{\ref*{fig:telpres-tree build}(a)}.
For each subsequent entry \((a, \text{seq}(a))\),
let $pref$ be the longest common prefix between $\text{seq}(a)$ and an existing path in the tree (compared by 2-tuples). 
Let $n$ be the node that contains the last 2-tuple of $pref$.
The construction proceeds according to one of the following three cases:

\begin{itemize}
   \item \textbf{Case 1: No match (\(|pref| = 0\)).} No common prefix exists. A new node containing the entire \(\text{seq}(a)\) is created and added as a child of the root node. For example, when inserting the first sequence \(\text{seq}(a_1)\) into an empty tree, a new node \(n_1\) is created as a child of the root to store the entire sequence (Fig.~\hyperref[fig:telpres-tree build]{\ref*{fig:telpres-tree build}(a)}).
    
    \item \textbf{Case 2: Full match (\(|pref| = |\textnormal{seq}(a)|\)).} The sequence \(\text{seq}(a)\) is a prefix of an existing path. No new nodes are needed. A new 3-tuple is added to the element table, where \(L(a)\) points to the last tuple in \(\text{seq}(a)\) within the sequence \(T\). For instance, \(\text{seq}(a_2)\) and \(\text{seq}(a_3)\) are prefixes of \(\text{seq}(a_1)\), so they share node \(n_1\) without modification (Fig.~\hyperref[fig:telpres-tree build]{\ref*{fig:telpres-tree build}(b)}).

    \item \textbf{Case 3: Partial match (\(0 < |pref| < |\textnormal{seq}(a)|\)).}
    \(\text{seq}(a)\) is split into two parts: the prefix \(pref\) and a remainder \(\text{seq}_{rem}(a)\). Two sub-cases arise:
    \begin{itemize}
        \item \(|pref| = |path(n_\text{root},n)|\): The prefix \(pref\) perfectly matches the path up to and including the entire sequence \(T\) in node \(n\). The handling of the remaining sequence \(\text{seq}_{rem}(a)\) is refined into two cases:
        \begin{itemize}
            \item If \(n\) has no children (is a leaf node), \(\text{seq}_{rem}(a)\) is appended directly to the sequence \(T\) in node \(n\), extending the existing path.
            \item If \(n\) already has children (is an internal node), a new child node containing \(\text{seq}_{rem}(a)\) is created and added under \(n\).
        \end{itemize}
        \item \(|pref| < |path(n_\text{root},n)|\): The prefix \(pref\) matches only a part of the sequence \(T\) in node \(n\). This requires splitting \(n\). The sequence \(T\) is divided into \(T^{pref}\) (the matching part) and \(T^{rem}\) (the remainder). The sequence in node \(n\) is truncated to \(T^{pref}\). Assuming the total number of nodes is i, two new nodes are then created as children of \(n\): (1) a node \(n_{i+1}\) containing \(T^{rem}\), which inherits all original children of \(n\); and (2) a node \(n_{i+2}\) containing \(\text{seq}_{rem}(a)\).
    \end{itemize}
For example, the prefix shared between $\text{seq}(a_4)$ and the path starting at \(n_1\) has a length of 2. This is shorter than both sequences, triggering a node split (Fig.~\hyperref[fig:telpres-tree build]{\ref*{fig:telpres-tree build}(c)}). Node \(n_1\) is truncated, and two new children are added: \(n_2\) (containing the remainder of the original \(T_1\)) and \(n_3\) (containing the remainder of \(\text{seq}(a_4)\)).

\end{itemize}

In all, the final constructed LFVT is shown in Fig.~\hyperref[fig:telpres-tree build]{\ref*{fig:telpres-tree build}(d)}. For the same $\text{S}$, the FVT requires 9 nodes, whereas the more compact LFVT requires only 5.
After the construction of the LFVT, the R-S Join computation using our proposed LFVT with length filter strategy, denoted as CF-RS-Join/LFVT, is similar to that of CF-RS-Join/FVT.

\section{Distributed CF-RS-Join with MapReduce}
\label{sec:FastAP-treeR-SJ}
To handle large-scale datasets, our algorithm is designed for parallel execution. 
For datasets that fit in a single machine's memory, the join process can be accelerated on multi-core architectures by partitioning \(\mathcal{R}\) to traverse the FVTs in parallel. 
When the data volume exceeds the memory capacity of a single machine, we employ a distributed computing approach using the MapReduce paradigm. 
This section presents our MapReduce-based approach, termed \textit{MR-CF-RS-Join/FVT}. We first discuss the key challenges and our proposed solutions and then detail the implementation.
These design principles also apply to CF-RS-Join/LFVT.

\subsection{Limitations of existing distributed approaches}
Most existing distributed R-S Join algorithms follow the filter-and-verification paradigm, typically requiring at least two MapReduce jobs. 
They first generate candidates using filters, then verify their similarity to obtain results.
The candidate generation phase often becomes the primary bottleneck, producing massive intermediate data that necessitates costly network shuffling and disk I/O.
More critically, it frequently leads to severe data skew, where uneven candidate distribution across partitions causes significant load imbalance during verification, severely degrading overall performance.

\begin{figure*}[tb]
    \centering
  \includegraphics[page=1,width=\linewidth]{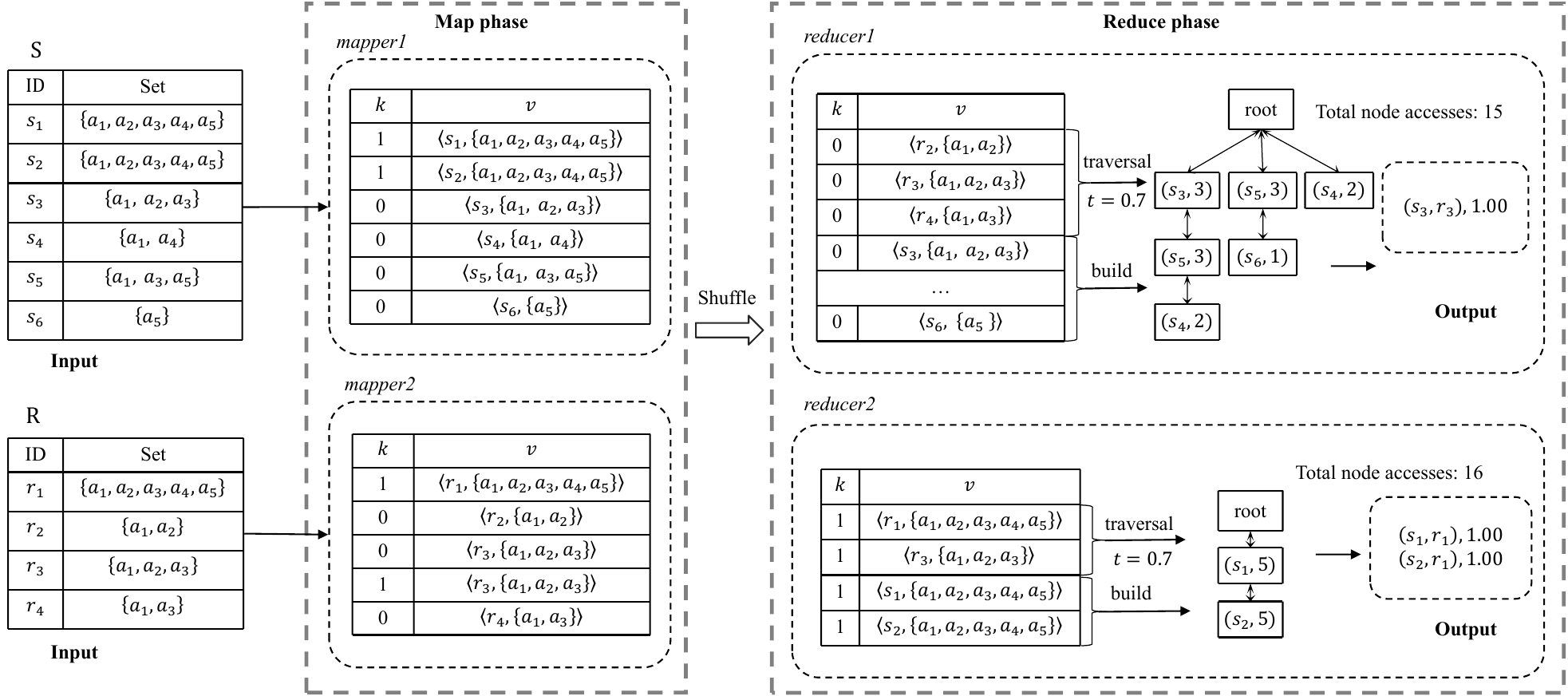}
   \caption{MR-CF-RS-Join/FVT over \(\text{R}\) and \(\text{S}\) $(k=2)$}
    \label{fig:FastAP-treeR-SJ}
\end{figure*}

\subsection{A single-stage MapReduce design}
Our \textit{MR-CF-RS-Join/FVT} algorithm overcomes these limitations by integrating filtering and verification into a candidate-free MapReduce job. 
This single-stage design provides its core efficiency advantage, 
the key is the FVT, which eliminates the need for an explicit candidate generation step.

The algorithm proceeds as follows. The Map phase partitions the data from both collections \(\mathcal{R}\) and \(\mathcal{S}\). 

To mitigate the load imbalance issue, we propose a novel load-aware data partitioning strategy. Inspired by the design of BundleJoin~\cite{yang2020distributed}, our approach aims to evenly distribute the computational workload across all reducers. In contrast, conventional hash-based~\cite{Le2014} or range-based~\cite{Myung2016} methods are often prone to severe data skew and consequent workload imbalances.

In the Reduce phase, each reducer first constructs an FVT from its assigned partition of \(\mathcal{S}\) sets. 
Then, it iterates through its partition of \(\mathcal{R}\) sets, directly probing each set against the FVT to efficiently discover all similar pairs without having generated any intermediate candidates. 
By eliminating the candidate generation phase, our algorithm avoids the massive I/O and data shuffling overheads inherent to existing approaches, resulting in substantially improved performance.

\subsection{Load-aware data partitioning strategy}
Our load-aware data partitioning strategy is executed in a driver program, which first
establishes a precise workload estimation model and then uses
dynamic programming to determine the optimal partitioning
scheme. The resulting length boundaries are then broadcast to
all mapper tasks to guide data partition and distribution in the subsequent
MapReduce job.

\subsubsection{Workload estimation}
To achieve balanced partitions, we first establish a cost model to estimate the workload of a single reducer. A reducer's workload is defined by the range of set size range \([\lambda_b, \lambda_e]\) it handles, where \(\lambda_b\) and \(\lambda_e\) represent the beginning (lower) and ending (upper) bounds, respectively. The computational load for a reducer assigned this range, denoted by \(\text{load}(\lambda_b, \lambda_e)\), consists of two main parts: (1) the cost of building an FVT from the assigned subset of \(\mathcal{S}\), and (2) the cost of probing this tree with the corresponding sets from \(\mathcal{R}\). The total load is formulated in Equation \eqref{equ:load}.

\begin{equation}\label{equ:load}
\footnotesize
\text{load}(\lambda_b,\lambda_e) = \left(\sum_{i=\lceil \lambda_b \times t \rceil}^{\lfloor \lambda_e/t \rfloor} i \cdot C_r(i)\right) \times \left(\sum_{j=\lambda_b}^{\lambda_e} j \cdot C_s(j)\right) + \sum_{x=\lambda_b}^{\lambda_e} x \cdot C_s(x)
\end{equation}
where $C_r(i)$ and  $C_s(i)$ denote the number of sets in $\mathcal{R}$ and $\mathcal{S}$ with length $i$, respectively.

\subsubsection{Optimal partitioning with dynamic programming}
Our goal is to partition the entire range of set size, \([l_{\min}, l_{\max}]\), into \(k\) disjoint sub-ranges, \(\{[\lambda_{b,1}, \lambda_{e,1}], \dots, [\lambda_{b,k}, \lambda_{e,k}]\}\), where \([\lambda_{b,i}, \lambda_{e,i}]\) is the set size range for the $i$-th reducer, minimizing the maximum load across reducers. We solve this optimization problem using dynamic programming. Let \(\psi(\lambda_b, \lambda_e, k)\) be the minimum possible maximum load when partitioning the length range \([\lambda_b, \lambda_e]\) across \(k\) reducers. The problem can be solved with the recurrence in Equation \eqref{equ:dp}.

\begin{equation}
\small
\label{equ:dp}
\psi(\lambda_b,\lambda_e,k)=
\begin{cases}
\text{load}(\lambda_b,\lambda_e), & k=1,\\[4pt]
\displaystyle \min\limits_{\lambda_b \le i < \lambda_e} \max \bigg(
\begin{aligned}
        &\psi(\lambda_b, i, k-1),\\
        &\text{load}(i+1,\lambda_e)
\end{aligned}
\bigg), & k\ge 2,\\[6pt]
0, & \text{otherwise}.\\
\end{cases}
\end{equation}    
By solving this recurrence for \(\psi(l_{\min}, l_{\max}, k)\), we obtain the optimal set of length boundaries for each reducer, which we denote as \(\{\langle i, [\lambda_{b,i}, \lambda_{e,i}]\rangle \mid 1 \leq i \leq k\}\).

\begin{example}
Fig. \ref{fig:FastAP-treeR-SJ} illustrates the computation process of the distributed CF-RS-Join/FVT over \(\text{R}\) and \(\text{S}\). With \(k=2\), we first compute the optimal partitioning strategy. The length range of \(\text{S}\) is \([1, 5]\). The dynamic programming solution yields a partition where \textit{reducer1} handles lengths \([4,5]\) and \textit{reducer2} handles \([1,3]\). Consequently, sets \(s_1, s_2\) are routed to \textit{reducer2}, while \(s_3, s_4, s_5, s_6\) go to \textit{reducer1}. Sets from \(\text{R}\) are routed accordingly based on potential length overlaps.
\end{example}

\subsubsection{Join execution}
Our load-aware data partitioning strategy is executed in a single MapReduce job, comprising the following Map and Reduce phases.
\begin{enumerate}
    \item \textbf{Map phase:} Mappers process records from \(\mathcal{R}\) and \(\mathcal{S}\) and emit key-value pairs, where the key is the target reducer ID.
    For each record \(s \in \mathcal{S}\), its length \(|s|\) falls into a specific partition \([\lambda_{b,i}, \lambda_{e,i}]\). The mapper emits this record to the corresponding \(reducer \, i\).
    For each record \(r \in \mathcal{R}\), its length \(|r|\) determines a range of potential matching lengths in \(\mathcal{S}\), i.e., \([\lceil|r| \times t\rceil, \lfloor|r|/t\rfloor]\). This range may overlap with multiple reducer partitions. Therefore, \(r\) is replicated and sent to all reducers whose length partitions overlap with this range. For instance, \(r_3\) (\(|r_3|=3\)) could potentially match sets of length \(\lceil 3 \times 0.7\rceil=3\) to \(\lfloor3/0.7\rfloor=4\). This length range \([3,4]\) overlaps with both reducer partitions \([1,3]\) and \([4,5]\), so \(r_3\) is sent to both reducers.
    \item \textbf{Reduce phase:} Each reducer \(reducer \, i\) receives a subset of \(\mathcal{S}\) and \(\mathcal{R}\). It first builds a local FVT using its assigned sets from \(\mathcal{S}\). Then, it probes the tree with sets from \(\mathcal{R}\) to perform the join and emit the final results. 
    \end{enumerate}

This load-aware data partitioning strategy ensures that the computational load is well-balanced across reducers (e.g., 15 vs. 16 accesses in our example, where an access represents a single comparison operation against the set size $|s|$ in a node).

\section{Performance Evaluation} \label{sec:Experimental Evaluation}
This section evaluates the performance of two proposed algorithms: CF-RS-Join and MR-CF-RS-Join. 
CF-RS-Join introduces a novel candidate-free approach integrating filtering and verification to reduce I/O. However, its scalability is inherently limited by memory capacity and single-machine processing power. To overcome this, we developed MR-CF-RS-Join, which extends the core logic to a distributed environment using MapReduce. To validate our design, we conduct two key experiments: (1) Analyzing the execution time speedup ratio and memory consumption of MR-CF-RS-Join relative to CF-RS-Join across various datasets to quantify the performance gains from distributed computing. (2) Benchmarking MR-CF-RS-Join against other SOTA distributed algorithms across a range of metrics, including execution time, scalability, memory usage, and disk I/O to demonstrate its competitiveness.

\begin{table*}[!htb]
\footnotesize
\caption{The details of $\mathcal{R}$ and $\mathcal{S}$ in seven real-world datasets}
\tabcolsep=0.13cm
    \begin{tabular}{ccrrrrrrrrrrrrrr}
    \toprule  
       Dataset &$|\mathcal{R}|$ or $|\mathcal{S}|$ & \multicolumn{3}{c}{$\text{avg},\min,\max |r|$} & $|\mathcal{R}'|$& \multicolumn{3}{c}{$\mathrm{avg},\min,\max$ $|\mathcal{\text{seq}}(a)|$ of $\mathcal{R}$} & \multicolumn{3}{c}{$\text{avg},\min,\max |s|$} & $|\mathcal{S}'|$& \multicolumn{3}{c}{$\mathrm{avg},\min,\max$ $|\mathcal{\text{seq}}(a)|$} of $\mathcal{S}$ \\  
    \midrule
 
 Dblp &500K & 15.555 & 1 & 203 & 275,006 &
    28.274 & 1 & 165,088 &  15.743 & 1 & 218 & 274,602 &
    28.658 & 1 & 168,989\\
    
     Kosarak & 500K & 11.589 & 1 & 2,497 & 35,673 &
          103.679 & 1& 214,606 & 11.545&1&2,490 &35,226&
          104.447& 1& 213,809\\
    LiveJ &1.5M&
        36.237&1&300&4,362,456& 11.998 &1 & 484,572 & 
        36.342 & 1 & 300 & 4,361,168 &12.034 & 1 & 484,470 \\ 

    Querylog &600K&
        1&1&1&600,000& 1 &1 & 1 &
        1&1&1&600,000& 1 &1 & 1 \\
    
     Enron&300K&141.604&1&3,162 &791,165& 25.398 &1 &112,745 &132.261&1&3,162 &737,771&  27.008 &1 &120,671\\

    Orkut &1.4M&
        120.022&1&40,426&7,229,601& 22.779 &1 & 1,372,131 & 
        120.274&1&14,193&7,228,835& 22.832 &1 & 1,372,288 \\
    Facebook & 297K &
        20.610&4&775&311,078& 19.709 &1 & 253,963 & 
        20.610&4&775&311,078& 19.709 &1 & 253,963 \\
      \bottomrule  
    
    \end{tabular}
    
    \label{tab:dataset details}
\end{table*}

\subsection{Experimental Setup}
Our empirical evaluation utilizes seven real-world datasets: Dblp~\cite{wang2012can}, Kosarak~\cite{kosarakandaccidents}, Enron~\cite{wang2012can}, LiveJ~\cite{mann2016empirical}, Orkut~\cite{mann2016empirical}, Querylog~\cite{wang2012can}, and Facebook~\cite{facebook2009}. To create the input collections, we randomly sample disjoint sets for \(\mathcal{R}\) and \(\mathcal{S}\) from each source and remove duplicates. For the smaller Facebook dataset, \(\mathcal{R}\) and \(\mathcal{S}\) are identical. The key characteristics of the resulting datasets are detailed in Table~\ref{tab:dataset details}.

\begin{figure}[tb]
    \centering
  \includegraphics[width=3.5in]{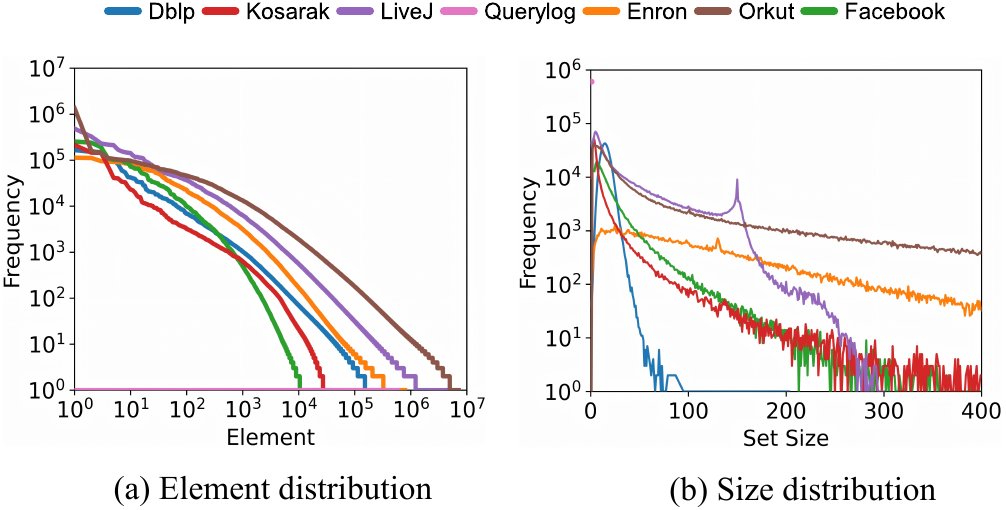}
   \caption{Histogram of elements and set sizes}
    \label{fig:distribution}
\end{figure}

These datasets were chosen to cover a wide spectrum of data distributions, as visualized in Figure~\ref{fig:distribution}. 
They differ in aspects such as element and set size distribution, and this diversity ensures that the performance of our algorithm can be fully evaluated under various conditions.
For example, LiveJ consists of many short sets from a large alphabet, whereas Enron and Orkut feature long sets. 
The element frequency in most datasets, like Orkut and LiveJ, roughly follows a Zipfian distribution, meaning there is a large number of infrequent elements (less than 10 occurrences).  
The set size distributions also differ significantly. We categorize a dataset as having a "narrow concentration range" if its set sizes are uniform (e.g., Dblp), and a "wide concentration range" if they are diverse (e.g., Enron, Orkut).

Apache Hadoop\footnote{Apache Hadoop, \url{https://hadoop.apache.org}} and Apache Spark\footnote{Apache Spark, \url{https://spark.apache.org}} are the two most prominent and widely-used implementations of the MapReduce framework. Their key distinction lies in intermediate data storage, where Hadoop persists intermediate results in the Hadoop Distributed File System (HDFS), whereas Spark caches them in memory. Consequently, Spark achieves higher efficiency for iterative algorithms with multi-stage data dependencies by reducing I/O overhead. As discussed in Sections \ref{sec:A-tree based method} and \ref{sec:FastAP-treeR-SJ}, CF-RS-Joins do not involve iterative processing with repeated data dependencies. Given that existing R-S Join algorithms are all Hadoop-based implementations, Hadoop is used to support programming and execution of the MapReduce based R-S Join baselines and our approaches. 

All experiments are conducted on a 17-machine (1 master and 16 slaves) cluster with Apache Hadoop 2.7 and OpenJDK 1.8.0.
The performance of CF-RS-Join's standalone version is evaluated on a single node. 
Each node is a CentOS 6.5 server with two Xeon E5-2680 2.8 GHz 10-core CPUs, 64GB of RAM, a 1TB hard disk, and a 1 Gbps Ethernet interconnect.
We optimize Hadoop resource allocation as shown in Table~\ref{tab:hadoop conf}. 
Since reducers need to buffer data, we allocate 4$\times$ more memory to reduce tasks than map tasks. By default, the number of reducers matches the number of nodes to avoid bottlenecks, as the slowest processing node determines overall execution time.
The execution time is measured using \texttt{java.lang.Date}, covering data input, index construction, search, verification, and data output.
For each algorithm, we measure and compare the number of reduce tasks run in parallel using multicore on each node and choose the optimal configuration.
Except for FS-Join, which runs two reduce tasks in parallel on each node on the Orkut dataset, other algorithms only run one reduce task.
Some algorithms experience timeouts when the threshold decreases because they fail to distribute the workload evenly, resulting in some nodes being overloaded.

\begin{table}[tb]
\footnotesize
\centering
 \caption{Hadoop configuration}
 \tabcolsep=0.05cm
    \begin{tabular}{lcc}
    \toprule
       Parameter  & Value & Description \\
    \midrule
        mapreduce.map.java.opts 
        & 4096MB & map task max memory\\  
        mapreduce.task.timeout & 5400000ms &
        task max execution time
        \\
      mapreduce.reduce.java.opts
        &20240MB & reduce task max memory\\
       yarn.scheduler.maximum-allocation-mb & 40000MB & job max memory\\
       yarn.scheduler.minimum-allocation-mb & 5120MB &job min memory \\
             \specialrule{0em}{1pt}{1pt}   
      yarn.nodemanager.resource.memory-mb &41440MB &
         \begin{tabular}[c]{@{}l@{}}node max available \\physical memory \end{tabular}\\
      \specialrule{0em}{1pt}{1pt}   
        \begin{tabular}[c]{@{}l@{}}
      yarn.nodemanager.resource.cpu-vcores
        \end{tabular} &20 &  \begin{tabular}[c]{@{}l@{}}
      YARN available\\virtual CPU count \\
        \end{tabular}\\
    \bottomrule
    \end{tabular}
    \label{tab:hadoop conf}
\end{table}

\begin{figure}[tb]
    \centering
  \includegraphics[width=3.5in]{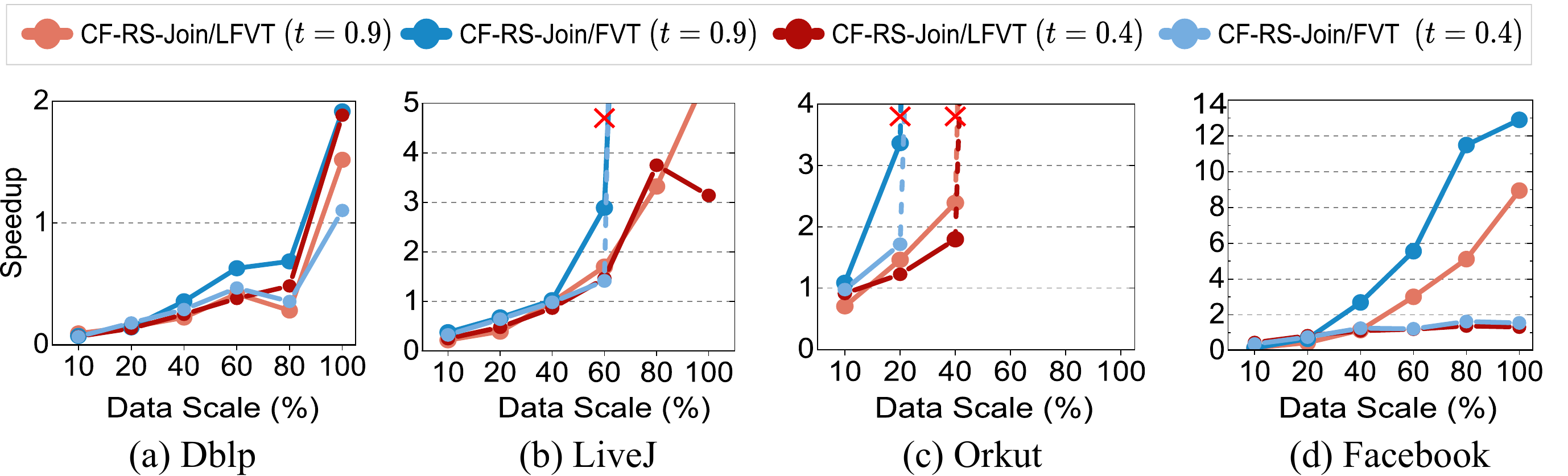}
   \caption{
   Speedup ratio of MR-CF-RS-Join vs. CF-RS-Join. 
   The \textbf{\textcolor{red}{$\times$}} represents out-of-memory.
   }
    \label{fig:speedup}
\end{figure}

\begin{figure*}[tb]
    \centering
  \includegraphics[width=6in]{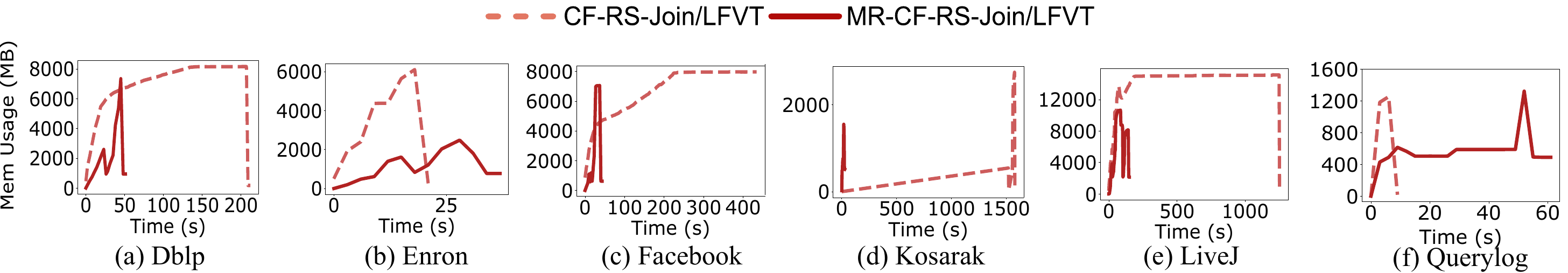}
   \caption{
   Memory usage of CF-RS-Join and MR-CF-RS-Join. The Orkut dataset is excluded due to timeout.}
    \label{fig:stand-mem}
\end{figure*}

\subsection{MR-CF-RS-Joins vs. CF-RS-Joins}
This section analyzes the scalability of MR-CF-RS-Join by comparing it with the standalone CF-RS-Join and evaluating their execution time speedup and memory consumption on various datasets.

First, we assess the performance in terms of execution time speedup.
As the data volume of four datasets increases from 10\% to
100\% at $t=0.9$ and $t=0.4$,
the speedup of MR-CF-RS-Join over CF-RS-Join becomes more pronounced, reaching up to 12.90$\times$ on the full Facebook dataset.  
For small datasets (e.g., Dblp and LiveJ), the speedup ratio may drop below 1.0 as inherent Hadoop overheads (e.g., starting/stopping jobs and transferring data between the cluster nodes) outweigh computational benefits, which aligns with prior findings \cite{fier2018set}.
Notably, the scalability limit of CF-RS-Join/FVT is evidenced by its out-of-memory (OOM) failures on larger datasets, such as the 80\%-scaled LiveJ ($t = 0.9$), 40\%-scaled Orkut ($t = 0.9$), and 80\%-scaled Orkut ($t = 0.4$), which further underscores MR-CF-RS-Join's performance advantage at data scale.

Second, we compare the memory usage.
Fig.~\ref{fig:stand-mem} illustrates the memory footprint of MR-CF-RS-Join/LFVT and CF-RS-Join/LFVT across six datasets at $t= 0.4$.
For MR-CF-RS-Join/LFVT, we report the
peak memory consumption observed on any single node across the cluster.
The results show a substantial reduction in memory footprint, often consuming less than half the memory of the standalone version. 
This efficiency is most critical for datasets with a "wide concentration range" like Enron and Orkut. For such data, the diverse set sizes force the CF-RS-Join to build a single, massive LFVT, exacerbating memory pressure and leading to the observed wider gap and OOM failures. 
MR-CF-RS-Join mitigates this bottleneck by enabling each reducer to construct a much smaller local LFVT. Consequently, while memory usage is comparable on datasets with a "narrow concentration range" (Dblp, Querylog), the dramatic reduction on more challenging datasets validates the effectiveness of our distributed design.

\subsection{MR-CF-RS-Joins vs. Distributed Baselines}
In this section, we benchmark MR-CF-RS-Join against SOTA distributed baseline algorithms to evaluate its performance in terms of execution time, scalability, and resource consumption (memory and disk I/O).

Existing filter-and-verification based distributed R-S Joins include FS-Join \cite{rong2017fast}, RP-PPJoin \cite{vernica2010efficient}, RP-PPJoin+Bitmap \cite{sandes2020bitmap}, and FSSJ~\cite{metwally2024similarity}. RP-PPJoin, one of the entire set filter-and-verification based algorithms, has been regarded as the winning algorithm concerning runtime and robustness w.r.t. various data characteristics \cite{fier2018set}. RP-PPJoin with an additional bitmap filter \cite{sandes2020bitmap}, denoted as {\em RP-PPJoin+Bitmap}, speeds up R-S Joins without sacrificing accurate results. FS-Join, one of the set segment filter-and-verification based algorithms, uses the vertical partitioning method to balance the inverted lists, ensuring element popularity is roughly equal across all computing nodes.
Additionally, the recently proposed FastTELP-SJ~\cite{feng2023FPtree} has demonstrated its effectiveness on self-join (i.e., $\mathcal{R}=\mathcal{S}$). To comprehensively evaluate CF-RS-Join's performance, we also design and implement its R-S Join variant. 
FSSJ~\cite{metwally2024similarity} exploits the skew in element popularity to avoid two costly operations: processing elements with high frequency and broadcasting the ordered elements to all the executors. The codes of baselines 
\cite{feng2023FPtree,rong2017fast,vernica2010efficient,sandes2020bitmap,metwally2024similarity} are not publicly available, we implement them from scratch in Java based on their original papers. 
We meticulously follow the descriptions and apply the optimal parameters reported in their original papers to guarantee a rigorous and equitable experimental setup.

\subsubsection{Execution time}
\label{subsec:runtime}
\begin{figure*}[tb]
    \centering
  \includegraphics[width=6in]{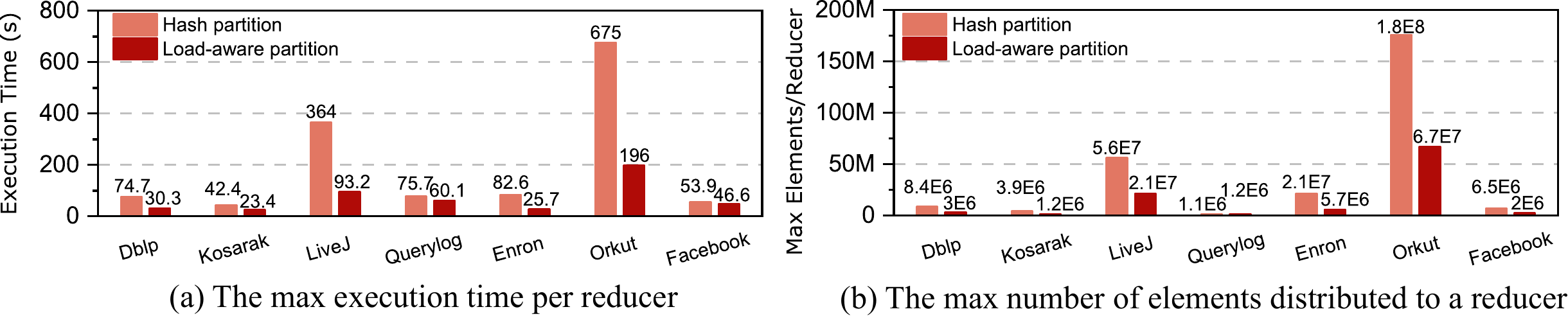}
   \caption{ 
   Ablation study of data partitioning strategy on distribution strategies $(t=0.9)$
   }
    \label{fig:parition}
\end{figure*}
We first evaluate our load-aware data partitioning strategy against the conventional hash-based partitioning (under hash-based partitioning, we construct a complete (L)FVT on each reducer and evenly distribute the $\mathcal{R}$ across $k$ reducers for R-S Join.).
Figs.~\hyperref[fig:parition]{\ref*{fig:parition}(a)} and \hyperref[fig:parition]{\ref*{fig:parition}(b)} show the execution time and the maximum number of elements assigned to any single reducer at $t=0.9$.
Our load-aware data partitioning strategy achieves an improvement in execution time of 68.9\%-74.4\% on datasets with wide concentration ranges (e.g., Enron, Orkut, LiveJ) and 13.5\%-44.8\% on those with narrow concentration ranges (e.g., Kosarak, Facebook).
For the Querylog dataset, where all sets are singletons, our strategy assigns all data to a single reducer. While this limits parallelism across reducers, it provides a net performance gain by avoiding the high I/O cost of replicating the $\mathcal{S}$ collection across all reducers, a step required by the hash-based strategy.

\begin{figure*}[tb]
    \centering
  \includegraphics[width=7in]{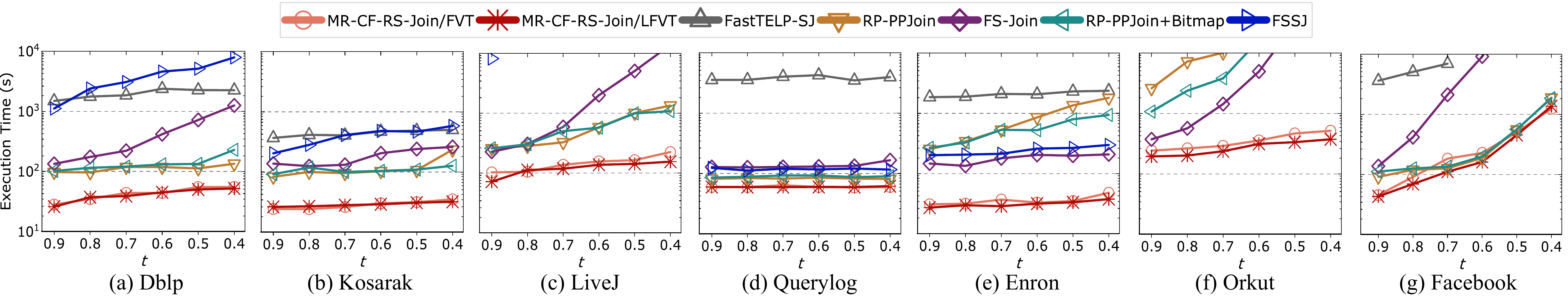}
   \caption{
   The execution time of distributed R-S Joins against the threshold
   }
    \label{fig:Runtime of distributed SOTA}
\end{figure*}

\begin{figure*}[tb]
    \centering
  \includegraphics[width=7in]{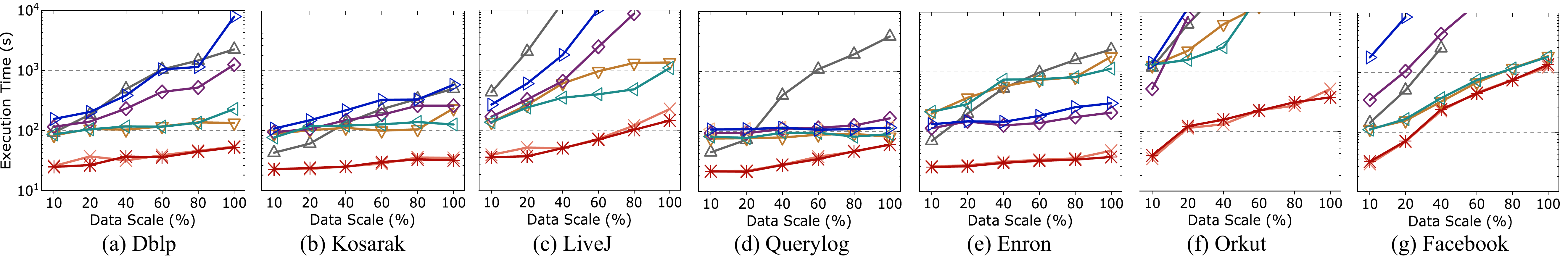}
   \caption{
   The execution time of distributed R-S Joins against the data scale $(t=0.4)$
   }
    \label{fig:scale of distributed SOTA}
\end{figure*}

\begin{figure}[tb]
    \centering
  \includegraphics[width=3.5in]{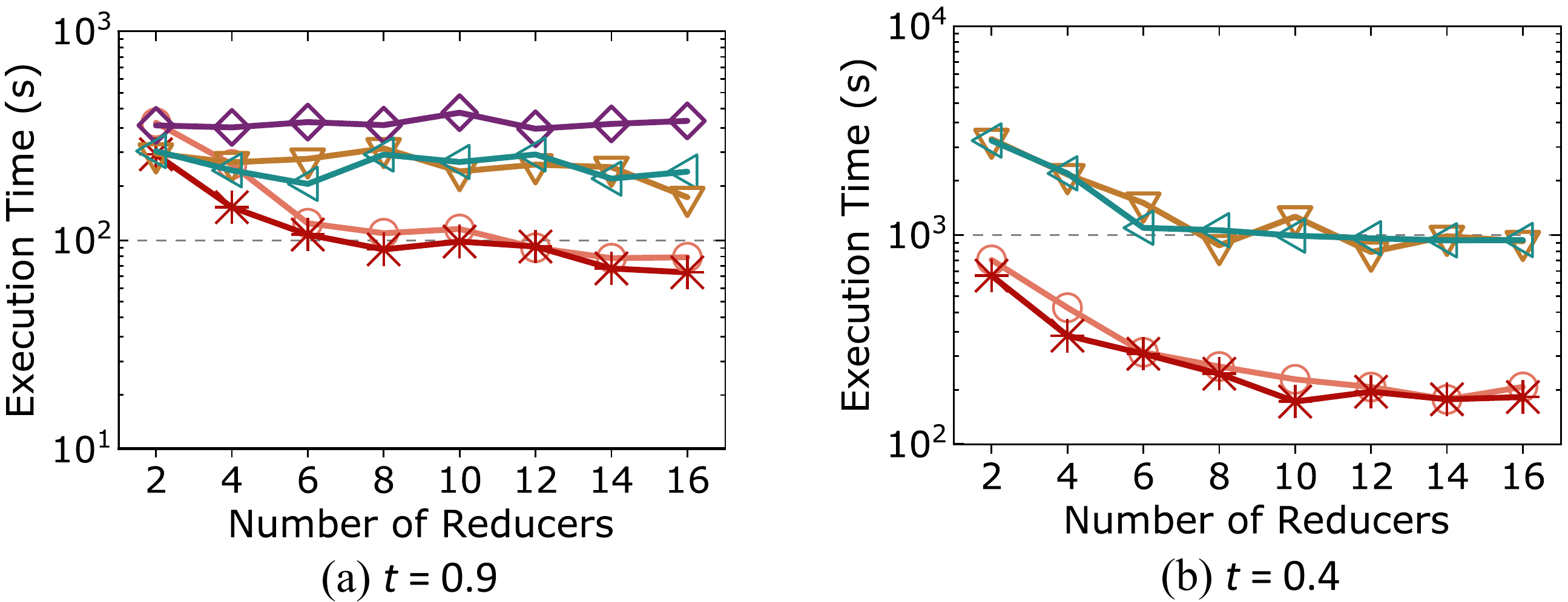}
   \caption{
   Execution time of distributed R-S Joins on the LiveJ dataset against cluster size (timeouts excluded).
   }
    \label{fig:scaleout}
\end{figure}
Fig.~\ref{fig:Runtime of distributed SOTA} illustrates the execution time of seven distributed algorithms across various thresholds. 
Our MR-CF-RS-Join consistently outperforms SOTA competitors on all datasets, which stems from three key designs: 
(1) an effective load-aware data partitioning strategy that balances computation in each reducer; 
(2) an in-memory, candidate-free join process that eliminates intermediate I/O overhead;
and (3) the accumulation of $|r \cap s|$ during tree traversal, enabling direct similarity computation in the verification stage.  
In contrast, FastTELP-SJ is often the slowest because it builds a large index on $\mathcal{R} \cup \mathcal{S}$ and employs inefficient sequential searches to locate child nodes in its tree structure, incurring substantial overhead.

On the Dblp and Kosarak datasets, the relative performance ranking remains consistent, with our algorithm always being the fastest.
On Kosarak at $t=0.9$, MR-CF-RS-Join/FVT is faster than RP-PPJoin, RP-PPJoin+Bitmp, FS-Join, FastTELP-SJ, and FSSJ by $2.20\times$, $2.54\times$, $4.30\times$, $13.18\times$, and $6.87 \times$, respectively.
At $t=0.4$, MR-CF-RS-Join/FVT is faster than RP-PPJoin, RP-PPJoin+Bitmap, FS-Join, FastTELP-SJ, and FSSJ by $7.74\times$, $5.59\times$, $9.05\times$, $18.30\times$, and $20.70\times$, respectively.

As $t$ decreases, the execution time of FS-Join and FSSJ changes the most.
On Dblp, for instance, FSSJ's execution time increases eightfold as $t$ drops from 0.9 to 0.4. This is because the set lengths in Dblp are clustered within $[10,25]$, rendering length-based filters ineffective,
causing FSSJ to generate and verify billions of candidate pairs—a costly process—even at $t=0.9$ and $t=0.8$ where no true similar pairs are present.

On the LiveJ dataset, the execution time of all competing algorithms degrades significantly with decreasing thresholds, whereas MR-CF-RS-Join remains stable.
On the Querylog dataset, all sets are singletons, rendering length-based filter strategies ineffective. Our algorithm maintains its advantage due to its single MapReduce job architecture, whereas multi-job competitors suffer from higher framework overhead.

On the Enron dataset, MR-CF-RS-Join significantly outperforms the runner-up FS-Join by an order of magnitude. 
Our algorithm often completes the entire join process before other methods finish their candidate generation phase.
Notably, as the threshold decreases from 0.7 to 0.4, the execution time gap between RP-PPJoin and RP-PPJoin+Bitmap widens significantly. 
This happens since the declining effectiveness of prefix, length, and suffix filter strategies in RP-PPJoin causes a 3.1$\times$ increase in candidates (from 1.9M to 7.8M).
The additional bitmap filter in RP-PPJoin+Bitmap, nevertheless, effectively prunes most of these candidates (filtered count rises from 1.8M to 6.0M), which mitigates the rise in its verification time.

On the Orkut dataset, our performance advantage grows, especially at lower thresholds.
FastTELP-SJ triggers the \texttt{mapreduce.task.timeout} limit during the Reduce phase due to the costly operations of constructing/traversing its large tree structure.
Meanwhile, the effectiveness of RP-PPJoin's length and suffix filters decreases as $t$ drops from 0.9 to 0.7 (filtering rates drop from 72.61\% to 10.97\% and 91.01\% to 40.14\%, respectively), causing a $74\times$ explosion in candidate pairs. 
Although the bitmap filter performs well at $t=0.7$, filtering 98\% of the candidates, the sheer volume of initial candidates still overwhelms the verification phase, degrading its overall performance.

On the Facebook dataset, the concentrated distribution of set sizes reduces the effectiveness of length and position filters, resulting in significantly increased execution time for all algorithms as the threshold decreases.
For example, when the $t$ decreases from 0.9 to 0.8, the number of candidates of RP-PPJoin+Bitmap grows by 4.84$\times$ (from 6,848,179 to 39,974,103), and that of FS-Join grows by 30.58$\times$ (from 303,738 to 9,591,946).
The reduce task of FastTELP-SJ approaches timeout at high thresholds because the elements of $\mathcal{R}$ and $\mathcal{S}$ in Facebook are identical. This makes the length filter strategy for reducing inverted lists completely ineffective in the first MapReduce task, resulting in excessively large inverted lists distributed to each reducer.

\subsubsection{Scalability}
We evaluate scalability from two perspectives: against increasing data volume (data scalability) and against an increasing number of compute nodes (cluster scalability).

For data scalability, we conduct experiments on data subsets ranging from 10\% to 100\% of the original datasets at $t=0.4$. 
As shown in Fig. \ref{fig:scale of distributed SOTA}, our algorithm's execution time scales best as data volume grows. 
This superior performance is attributable to two main factors. 
First, our load-aware data partitioning strategy effectively balances workloads across nodes, thereby mitigating the issue of node overload that degrades the performance of competitors such as RP-PPJoin(+Bitmap).
This problem becomes particularly pronounced as the frequency of shared prefix elements rises with increasing data scale.
Second, our single-stage filter-and-verification achieves two key optimizations: (1) it completely eliminates I/O overhead from writing and reading candidates; (2) it directly utilizes the $|r \cap s|$ values computed during the search phase for verification.
This contrasts with other baselines, which suffer from escalating I/O costs and lack support for fast verification processes as data volumes increase. For example, RP-PPJoin(+Bitmap) requires traversing the elements of candidates to compute intersections, FSJoin must merge multiple candidate intersections before verification, and FSSJ needs to access quasi-suffixes during verification to calculate similarity.

For cluster scalability, we evaluate performance on the LiveJ dataset by scaling the number of nodes from 2 to 16. As shown in Fig. \ref{fig:scaleout}, MR-CF-RS-Join demonstrates strong scale-out capabilities. 
When the number of reducers is increased $8\times$, our algorithm achieves significant speedups: for the MR-CF-RS-Join/FVT, up to 5.57$\times$ ($t=0.9$) and 3.72$\times$ ($t=0.4$); and for the MR-CF-RS-Join/LFVT, 3.67$\times$ ($t=0.9$) and 3.37$\times$ ($t=0.4$). 
Notably, our approach maintains effective parallelization across more nodes than competing algorithms, indicating a more efficient use of distributed resources before performance gains diminish.

\subsubsection{Memory Usage}
\label{distributed-peakmemory}

\begin{figure*}[tb]
    \centering
  \includegraphics[width=7in]{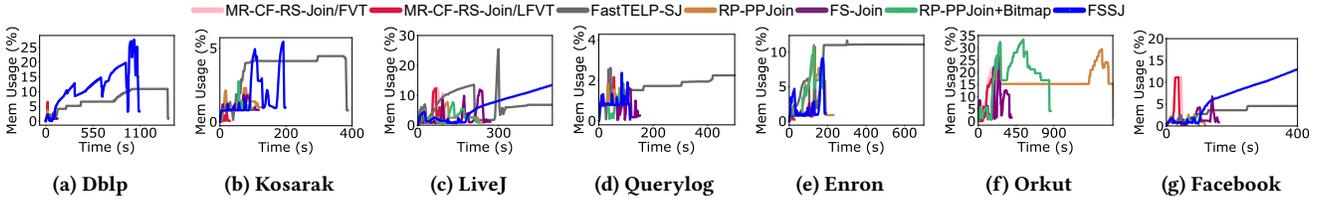}
   \caption{
   The memory usage of distributed R-S Joins $(t=0.9)$
   }
    \label{fig:menory usage}
\end{figure*}

\begin{figure*}[tb]
    \centering
  \includegraphics[width=7in]{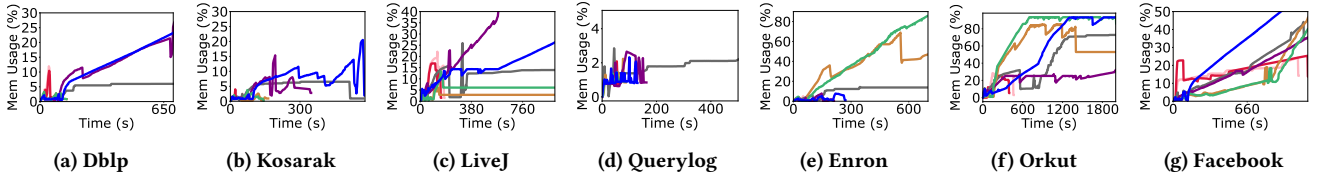}
   \caption{
   The memory usage of distributed R-S Joins $(t=0.4)$
   }
    \label{fig:menory usage2}
\end{figure*}

We measure runtime memory usage by recording the highest memory usage among all compute nodes at 3-second intervals. Each sampled value of used memory (obtained via \texttt{free -m}) is divided by 64GB
to calculate the memory usage rate. Crucially, we subtract the pre-execution \texttt{free -m} baseline from all measurements to isolate algorithm-specific memory usage from the operating system's inherent memory occupation. 

Although MR-CF-RS-Join is designed for in-memory computation, the memory usage rates are within an acceptable range.
At a high threshold of $t=0.9$ (Fig. \ref{fig:menory usage}), RP-PPJoin(+Bitmap) consumes minimal memory on most datasets (Dblp, LiveJ, Querylog, and Facebook). 
They employ the prefix filter to reduce index size and the number of candidates, e.g., filter approximately 99.6\% of elements in the Facebook dataset. 
On most datasets, RP-PPJoin+Bitmap exhibits higher memory consumption than RP-PPJoin due to its requirement of constructing and maintaining bitmap structures for each processed set during computation.
For FastTELP-SJ, the benefits gained from the filter strategy do not offset the overhead of 
building a single large tree
based on $\mathcal{R} \cup \mathcal{S}$.

At a lower threshold of $t=0.4$ (Fig. \ref{fig:menory usage2}),
our algorithm's memory consumption remains
low and stable as candidate-based methods struggle with exploding intermediate results.
For instance,
FS-Join exhibits a surge in memory usage on the Dblp and LiveJ datasets due to the substantial growth in candidates at lower thresholds (e.g., on LiveJ, the number of candidates increases from $10^9$ to $10^{10}$ when $t$ drops from 0.7 to 0.6).

\begin{table}[tb]
  \centering
  \caption{Disk usage of distributed R-S Join algorithms (MB).
  \textbf{Bold} and \underline{underline} values indicate the least and second least disk usage, respectively. "-" indicates the algorithm timed out.}
  \label{tab:disk usage}%
   \setlength\tabcolsep{2.5pt}
    \begin{tabular}{cccccccc}
    \toprule
    \textbf{$t$} & \textbf{Dataset} & 
     \textbf{MR-CF}
    & \textbf{PP} & \textbf{PP-BF}
    & \textbf{FastTELP} & \textbf{FS} & \textbf{FSSJ}\\
    \midrule

    \multirow{6}*{\makecell{0.9}}  
    &Dblp& \textbf{637} &  \underline{664} & 688 & 1851& 2078 & 915\\
    &Kosarak  & \textbf{121} & 878 & 890 & 827&1284& \underline{386}\\
    &LiveJ & \textbf{4419}  & 13129 & 13264& - & 9394 &\underline{6527}\\
    &Querylog & \textbf{49} & 176  & 186 &82 &2279 & \underline{64}\\
    
    & Enron &  \underline{1086}  & 23021 & 23075 & 4673 & \textbf{1042} & 2242\\
    &Orkut& \underline{12591}& 229884 & 230444 & - & \textbf{12001} & -\\
    &Facebook& \textbf{337} &\underline{997} & 1014 & 1444 & 2712 & -\\
    \midrule

    \multirow{6}*{\makecell{0.4}}  
    &Dblp & 3087 & 3005 & 3120 & \underline{1856}  & 143862 & \textbf{1052}\\
    &Kosarak  &\underline{521} & 4980 & 5053 &829 & 22148& \textbf{426}\\
    &LiveJ &\textbf{14636}& \underline{85426} & 86360 & - & 1046873 & -\\
    &Querylog & \textbf{49} & 176  & 186 &82 & 2279 & \underline{64}\\
    
    & Enron & \underline{3954} & 164661  & 165072 &  4698 & 4695 & \textbf{2308}\\
    &Orkut& \textbf{39538} & - & - & - & - & -\\
    &Facebook & \textbf{1326} & \underline{67928} & 68042 & - & - & -\\
    \bottomrule
    \end{tabular}%
\end{table}%

\subsubsection{Disk Usage}

Table \ref{tab:disk usage} details the disk usage of distributed algorithms, measured via Hadoop's built-in counter (FileSystemCounters) to track data written by map tasks. 
Since the intermediate data output by mappers is written to local disk, its volume directly determines the disk and net I/O overhead in the shuffle stage.
Our MR-CF-RS-Join is designed to minimize this metric by avoiding the generation of intermediate candidates—a primary source of disk I/O in distributed R-S Joins. 
During the Map phase, our algorithm only writes the original sets and their target reducer IDs, leading to a consistently low disk footprint.
Consequently, both MR-CF-RS-Join/FVT and MR-CF-RS-Join/LFVT, which share the same load-aware data partitioning strategy and thus distribute identical data volumes, exhibit identical disk footprints.

At $t=0.9$, MR-CF-RS-Join has the lowest disk usage on the Dblp, Kosarak, LiveJ, Querylog, and Facebook datasets, and the second
lowest on the Enron and Orkut datasets, being slightly higher than the FS-Join by less than 5\%. 

At $t=0.4$, MR-CF-RS-Join has the lowest disk usage on LiveJ, Orkut, Querylog, and Facebook datasets, and the second lowest on Enron and Kosarak datasets.
FastTELP-SJ has more disk usage than ours because its upper bound for element distribution volume during the Map phase reaches $((|\mathcal{S}| \times \overline{|{S}|} + |\mathcal{R}| \times \overline{|{R}|}) \times (k + 1) )$ elements, whereas ours remains at $(|\mathcal{S}| \times \overline{|{S}|} + |\mathcal{R}| \times \overline{|{R}|} \times k)$ elements. However, FastTELP-SJ employs an effective length-based filter strategy to reduce data distribution volume (when the $t$ decreases from 0.9 to 0.4, the data distribution volume in the Map phase only increases by 1\%-2\%), maintaining modest growth in disk usage when the $t$ decreases.
As the filtering effectiveness of methods like FS-Join and RP-PPJoin diminishes, they suffer from an explosion in intermediate candidates that must be written to disk. 
As shown in Table~\ref{tab:disk usage}, when the $t$ changes from 0.9 to 0.4, the disk usage of FS-Join on the Dblp, Kosarak, and LiveJ datasets expands by 68.23$\times$, 16.25$\times$, and 110.44$\times$, respectively. RP-PPJoin's disk usage expands by 3.53$\times$, 4.67$\times$, 5.51$\times$, 6.15$\times$, and 67.13$\times$ on Dblp, Kosarak, LiveJ, Enron, and Facebook datasets, respectively.
FSSJ generates fewer candidates by broadcasting quasi-prefix elements (elements with the lowest frequency), thereby reducing disk usage.
As $t$ decreases, some higher-frequency elements are added to the quasi-prefix and broadcast, increasing disk usage slightly (e.g., FSSJ's disk usage only increases by 14.97\%, 10.36\%, and 2.94\% on the Dblp, Kosarak, and Enron datasets).

\section{Conclusion}
\label{sec:conclusion}
Exact R-S Join using distributed architectures is needed in various applications like financial fraud detection, data mining, and large-scale genomic and metagenomic analysis.  
However, existing distributed R-S Join algorithms, built on the filter-and-verification framework, face a critical performance bottleneck when processing large-scale datasets: the generation of massive intermediate candidates. This leads to excessive disk I/O, network traffic, and pairwise verification costs, degrading performance.
To address this limitation, this paper introduces a novel candidate-free paradigm for exact R-S Joins. By proposing the Filter-and-Verification Tree (FVT) and its more compact variant, the Linear FVT (LFVT), we integrate the filtering and verification stages into a single, in-memory operation. This design fundamentally eliminates the generation of candidates.
We propose MR-CF-RS-Join/FVT and MR-CF-RS-Join/LFVT to exploit MapReduce for further speeding up the FVT based and LFVT based CF-RS-Join computation, respectively.
Experimental results demonstrate that our proposed algorithm using MapReduce, i.e., MR-CF-RS-Join/LFVT, achieves the best performance in terms of execution time, scalability, memory usage, and disk usage.

\section*{Acknowledgments}
The authors would like to thank the anonymous reviewers for their 
valuable comments and constructive suggestions, which significantly 
improved the quality of this paper.


 
\bibliographystyle{IEEEtran}
\bibliography{main}           

\newpage

 

\begin{IEEEbiography}[{\includegraphics[width=1in,height=1.25in,clip,keepaspectratio]{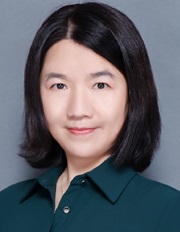}}]{Yuhong Feng} received her B.S. and PhD from University of Science and Technology of China and Singapore Nanyang Technological University
respectively. She is an associate professor at College of Computer Science
and Software Engineering, Shenzhen University, China. She was an assistant
researcher at SIAT, CAS and a postdoc fellow at HKPolyU. Her research interests include Parallel and Distributed Computing, Cloud computing, Software Security Analysis, and Smart Farming.
\end{IEEEbiography}

\begin{IEEEbiography}[{\includegraphics[width=1in,height=1.25in,clip,keepaspectratio]{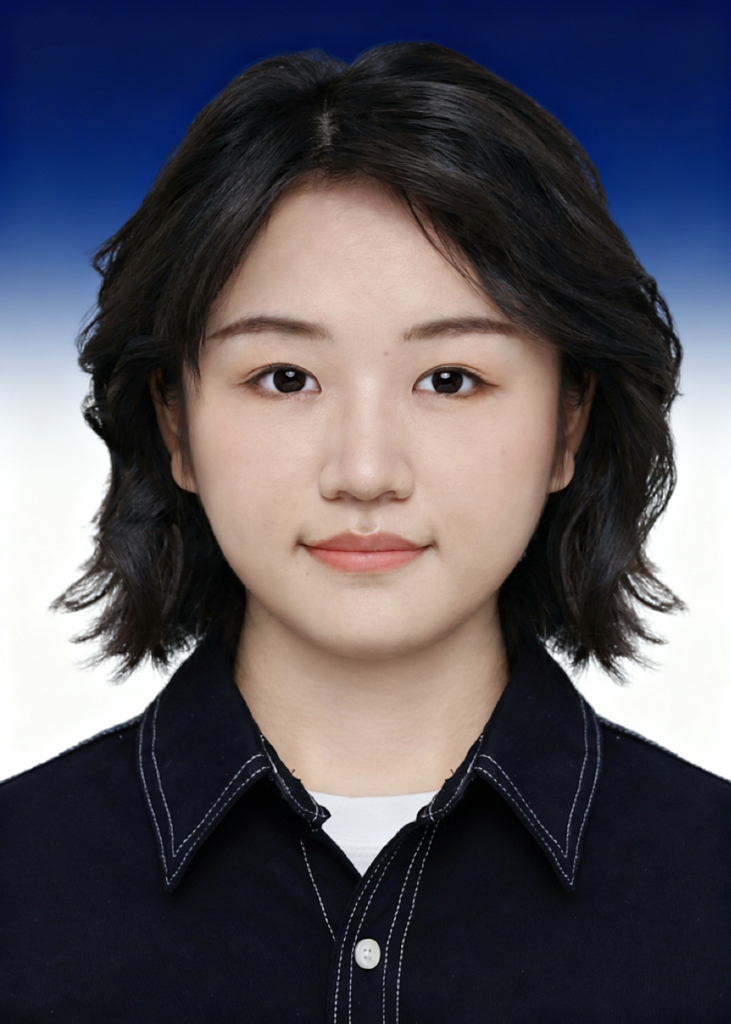}}]{Fangcao Jian}
is currently an M.S. candidate at the College of Computer Science and Software Engineering, Shenzhen University, China. She received her Bachelor’s degree from North China University of Science and Technology.
Her research interests include parallel and distributed computing and big data.
\end{IEEEbiography}

\begin{IEEEbiography}[{\includegraphics[width=1in,height=1.25in,clip,keepaspectratio]{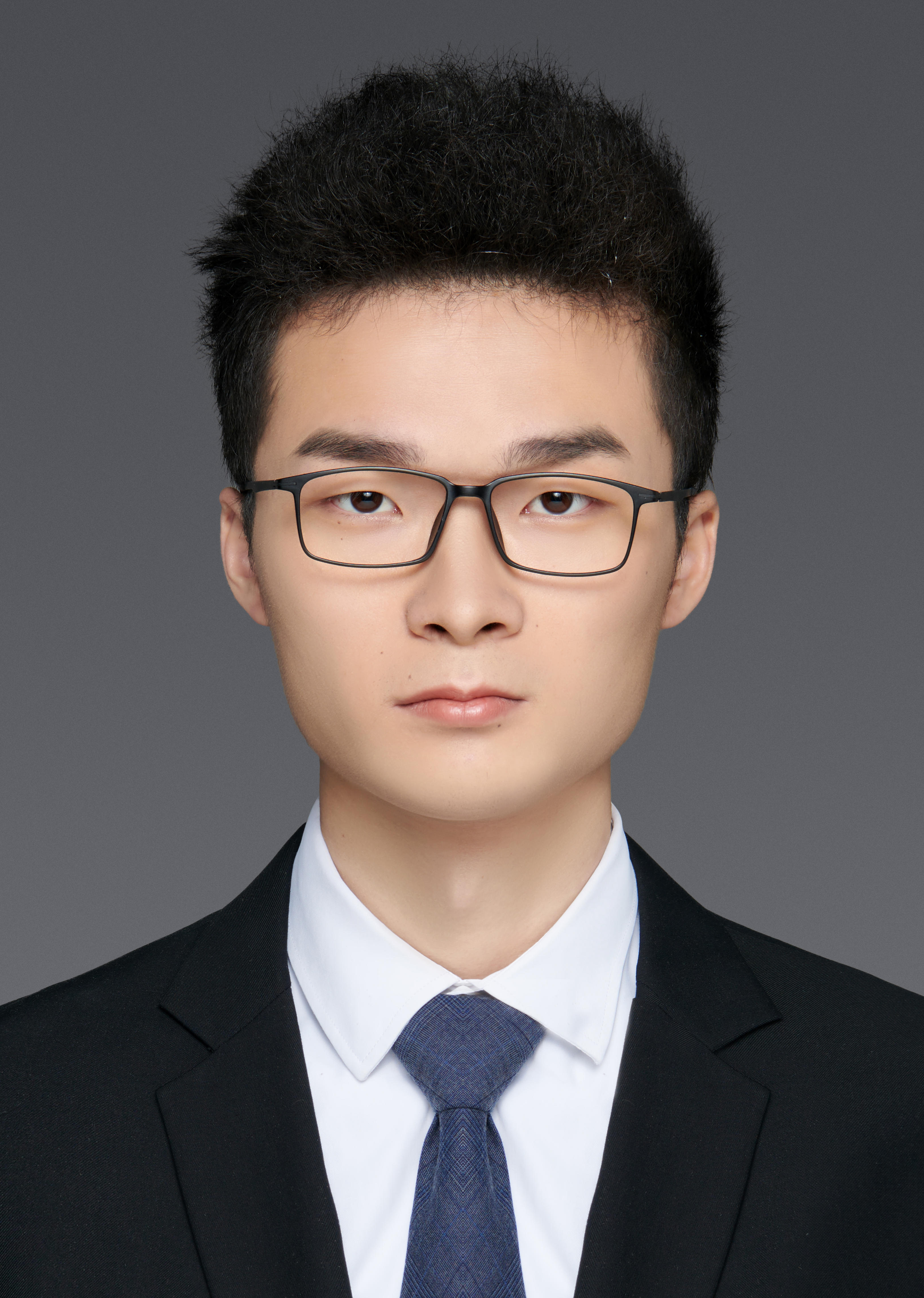}}]{Yixuan Cao} received his B.Eng. degree with honors from the College of Computer Science and Software Engineering, Shenzhen University, Shenzhen, China, where he is currently an M.S. candidate. His research interests primarily include software engineering, parallel computing, and artificial intelligence. He is an active open-source community contributor and has successfully merged multiple patches into the Linux kernel and LLVM.
\end{IEEEbiography}

\begin{IEEEbiography}[{\includegraphics[width=1in,height=1.25in,clip,keepaspectratio]{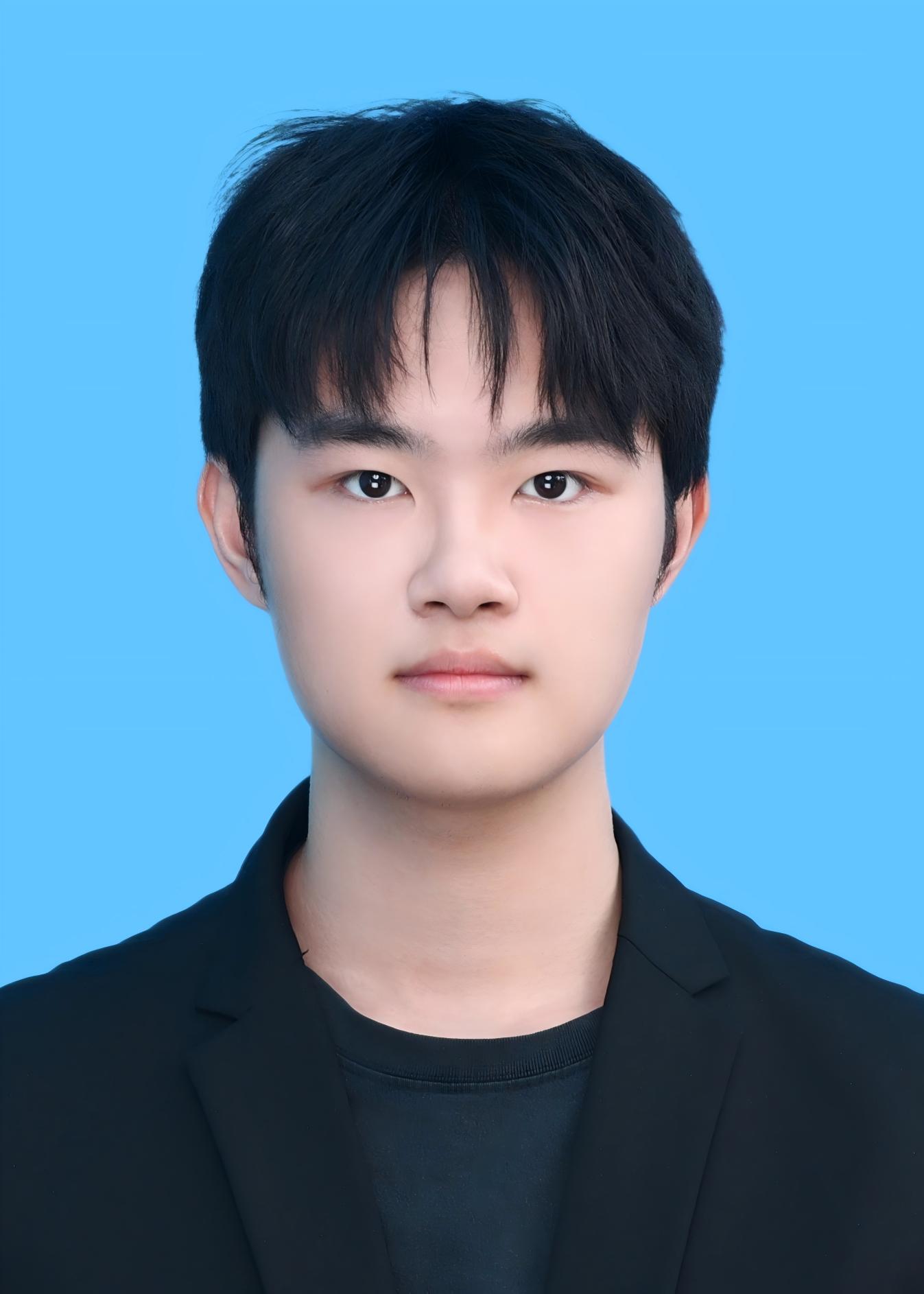}}]{Xiaobin Jian} is currently a fourth-year undergraduate student at the College of Computer Science and Software Engineering, Shenzhen University. He is soon to commence his Academic Master's program at the School of Computer Science and Innovation, Fudan University. During his Master's studies, he will focus on the cutting-edge research area of Generative Recommendation. 
\end{IEEEbiography}

\begin{IEEEbiography}[{\includegraphics[width=1in,height=1.25in,clip,keepaspectratio]{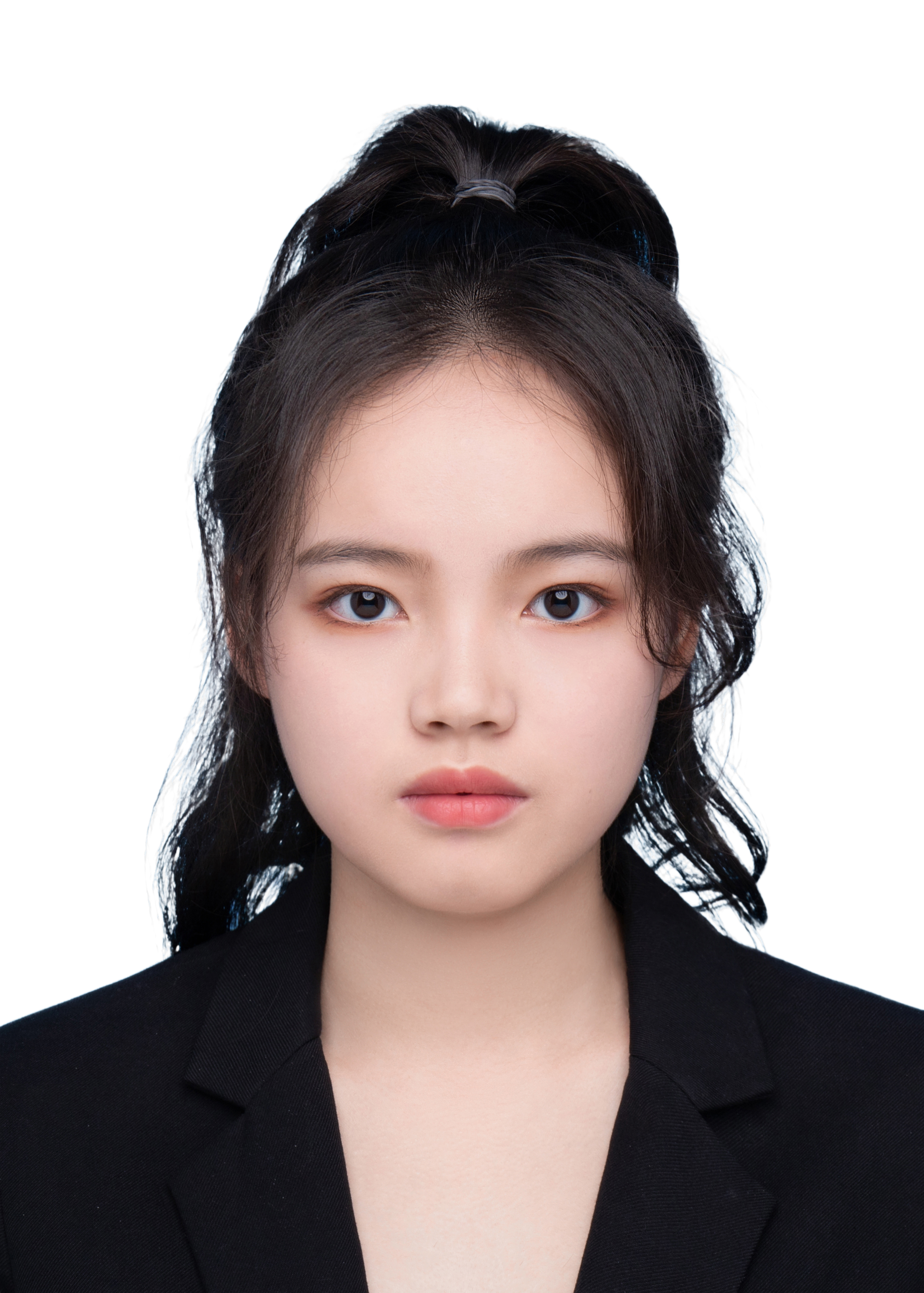}}]{Jia Wang} is a third-year undergraduate student at Shenzhen University, where she is enrolled in the School of Computer Science and Software Engineering, majoring in Software Engineering. Beyond her coursework, she is actively engaged in academic exploration, with her primary research interest centered on Parallel and Distributed Computing. She delves into cutting-edge topics such as high-performance computing and distributed system architecture.
\end{IEEEbiography}

\begin{IEEEbiography}[{\includegraphics[width=1in,height=1.25in,clip,keepaspectratio]{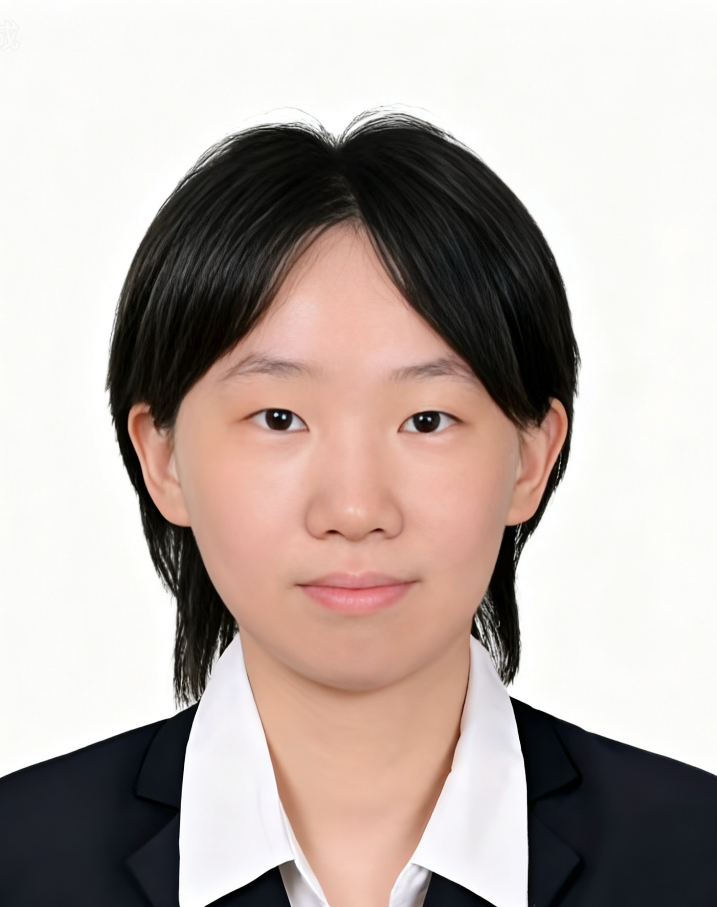}}]{Haiyue Feng} 
received the ME degree in computer technology from Shenzhen University, China, in 2025. She is currently working toward the PhD degree with the College of Computing and Data Science, Nanyang
Technological University, Singapore. Her research
interests include AI for education and Human-AI Interaction.
\end{IEEEbiography}

\begin{IEEEbiography}[{\includegraphics[width=1in,height=1.25in,clip,keepaspectratio]{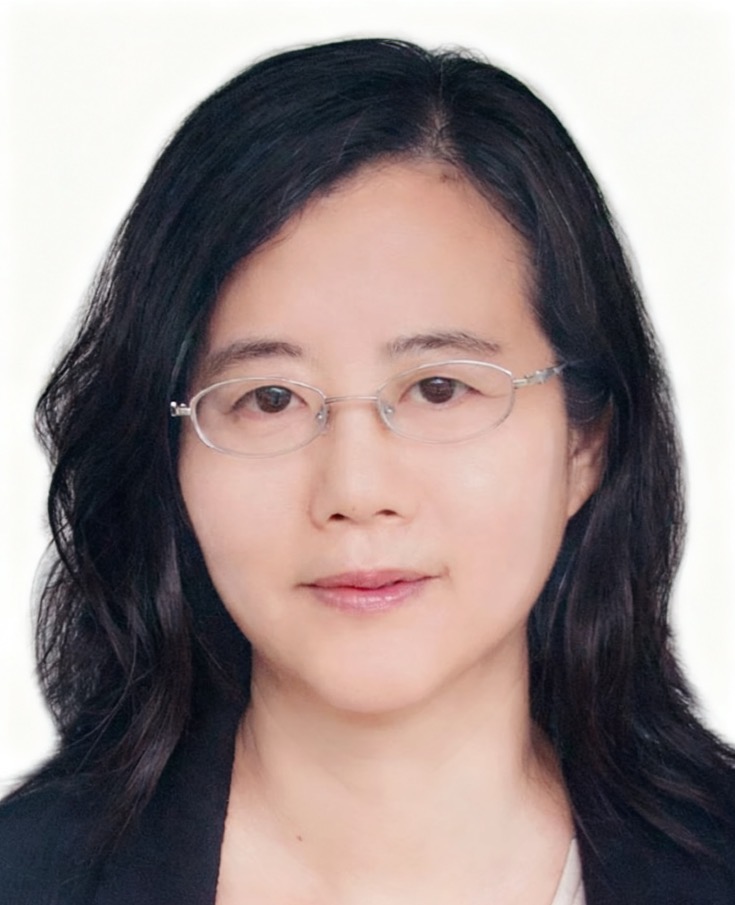}}]{Chunyan Miao (IEEE Fellow)} is a President’s Chair Professor and Chair of the College of Computing and Data Science at Nanyang Technological University (NTU), Singapore. She received her PhD from NTU and was an NSERC Postdoctoral Fellow at Simon Fraser University. Her research focuses on AI and its real-world applications in health, aging, education, and smart services. She has founded and leads key research centers, including the Joint NTU-UBC Research Centre of Excellence in Active Living for the Elderly (LILY) — Singapore’s first AI-driven center tackling aging — and the Alibaba-NTU Singapore Joint Research Institute, Alibaba’s first and largest such institute outside China. She serves as editor/associate editor for journals such as IEEE Internet of Things Journal and IEEE Access, and has chaired or served on program committees for conferences including ACM KDD and IEEE ICA. She also chairs the SCS AI Ethics Review Committee and participates in various national committees.
\end{IEEEbiography}



\vfill

\end{document}